\documentstyle[12pt,epsf]{article}
\begin{document}
 
\newcommand{\be}{\begin{eqnarray}}
\newcommand{\ee}{\end{eqnarray}}
\begin{flushright}
LBNL-47715
\end{flushright}
\vspace{3mm} 
\begin{center}
 
{\Large
{Asymmetries Between Strange and Antistrange Particle Production
in Pion-Proton Interactions\footnote{
This work was supported in part by the Director, Office of Energy
Research, Division of Nuclear Physics of the Office of High Energy
and Nuclear Physics of the U. S. Department of Energy under Contract
Number DE-AC03-76SF00098.}
}} \\[8ex]

T.D. Gutierrez$^a$ and R. Vogt$^{a,b}$\\[2ex]
 $^a$Physics Department\\
 University of California at Davis\\
 Davis, California\quad 95616  \\
 and\\
 $^b$Nuclear Science Division\\
 Lawrence Berkeley National Laboratory\\
 Berkeley, California\quad 94720 \\[6ex]

{\bf ABSTRACT}
\end{center}
\begin{quote}
Recent measurements of the asymmetries between Feynman $x$ distributions
of strange and antistrange hadrons in $\pi^- A$ interactions
show a strong effect as a function of
$x_F$.  We calculate strange
hadron production in the context of the intrinsic model and make predictions
for particle/antiparticle asymmetries in these
interactions. 
\end{quote}
\vspace{2mm}
\begin{center}
\end{center}
\newpage

\section{Introduction}

Flavor correlations
between the final-state hadron and the projectile have been observed 
in charm hadroproduction.  A strong leading particle effect was seen in
the difference between the $D^-$ and $D^+$ 
distributions at large Feynman $x$, $x_F = p_{||}/p_{\rm max}$, 
with pion projectiles \cite{E791,ddata1,ddata2,ddata3,ddata4,ddata5}.  
More recently, hyperon beams have been used to study charm
baryon distributions at high $x_F$ \cite{wa891,wa892,SELEX1,SELEX2}.  
Several of these experiments 
\cite{wa891,wa892,SELEX1,SELEX2,E791lam} have also studied the $x_F$-dependent
asymmetry between charm and anticharm baryons.  This asymmetry is 
defined as the ratio of, for example, the
difference between the $\Lambda_c$ and $\overline \Lambda_c$ 
$x_F$ distributions divided by 
their sums.  Only recently has such data become available
in the strange sector \cite{anjosmex1,anjosmex2}.

The strange/antistrange baryon asymmetries $A_\Lambda$, $A_{\Xi^-}$, and 
$A_\Omega$ have been measured in $\pi^-$-induced 
interactions at 500 GeV by the E791 Collaboration \cite{anjosmex1,anjosmex2}.  
The measurements are around $|x_F|<0.1$.  They find
that for $x_F > 0$, $A_\Lambda$ 
and $A_\Omega$ are nearly independent
of $x_F$ while $A_{\Xi^-}$ increases with $x_F$.
On the other hand, at negative $x_F$, only $A_\Omega$ is independent
of $x_F$.  The other asymmetries increase as $x_F$ decreases with $A_\Lambda
> A_{\Xi^-}$.
These measurements are inconsistent with PYTHIA \cite{PYT}
which produces essentially no asymmetry at forward $x_F$ while at
negative $x_F$, only $A_\Lambda$ is increasing.  
The trends of the data are consistent with qualitative expectations from
recombination models \cite{anjosmex1,anjosmex2}.

One such model that involves recombination with valence quarks was first 
developed to explain large $x$ production of charm in the proton structure
function, the ``intrinsic
charm'' model originally motivated in Refs.~\cite{intc1,intc2}.  
The model, including
leading-twist $c \overline c$ production was 
extended to charm hadron asymmetries such as $A_{D^-}$ in subsequent works 
\cite{VB,VBlam,GutVogt1}.  In this picture, the 
projectile can fluctuate into a Fock
state configuration with at least one $c \overline c$ pair as well as other
light $q \overline q$ pairs.  These charm quarks are comoving with the other
partons in the Fock state and thus can combine with these comoving partons
to produce charm hadrons at large $x_F$. 
The probability that the projectile 
fluctuates into a state with the projectile valence quarks and a
$c \overline c$ pair is $\approx 0.3$\% \cite{hsv}.  
The probability for intrinsic states with other $Q \overline Q$ pairs
scales as the square of
the constituent quark mass.  Since strange quarks are
lighter than charm quarks, the corresponding probability for intrinsic
$s \overline s$ pairs in the wavefunction should be 
significantly larger.

In this paper, we apply the combined leading-twist/intrinsic model of
Ref.~\cite{GutVogt1} to strangeness production.  We describe how we calculate
strangeness production at leading twist in Section 2.  Section 3 is devoted to
a description of the intrinsic model for strange quarks.  
Section 4 presents the calculations of strange hadron production asymmetries
in $\pi^- p$ interactions for both positive and negative $x_F$, 
the pion and proton fragmentation regions respectively.  
Our results are summarized in Section 5.

\section{Leading-Twist Strangeness Production}

The two component model we have used to study charm hadroproduction consists
of a perturbative, leading-twist, component that normalizes the cross section
at $x_F \sim 0$ and the intrinsic, higher-twist, component
\cite{VB,VBlam,GutVogt1}. When addressing strangeness production, however, 
we must keep in mind that quarks lighter than charm are 
difficult to treat within the context of perturbative QCD.  We
thus choose a set of parton distribution functions that is most compatible
with our needs and assume that the strange quark is massive, considering only 
the $gg \rightarrow s \overline s$ and $q
\overline q \rightarrow s \overline s$ production channels. 

Our treatment of strange quarks
as heavy is rather uncertain since strange quarks are 
considerably lighter than charm, $m_s \approx 150-500 \, {\rm MeV} \, \approx
(1/3-1/10) m_c$, and 
charm production is already subject to large corrections
beyond leading order \cite{smithv}.  The lower end of the
strange quark mass range is 
close to the $n_f = 3$ value of $\Lambda_{\rm QCD}$ and less than
the initial scale of all parton distribution functions.  The
strong coupling constant will thus be large and the leading order (LO)
cross section will only be a fraction of the complete result.  Therefore a 
perturbative treatment of strange quark production is dubious.

The GRV 94 LO proton
parton distribution functions \cite{GRV94} are most suitable
because the strange quark distribution vanishes at 
the initial scale $Q_0^2 = 0.4$ GeV$^2$.
The older GRV LO parton distribution functions \cite{GRV} employ a lower 
initial scale but an isospin symmetric sea ($\overline u = \overline d$) while
the most recent GRV 98 LO set \cite{GRV98} assumes a higher initial scale,
$Q_0^2 = 0.8$ GeV$^2$.  All other recent parton distribution functions 
\cite{PDFLIB} employ a scale of 1 GeV$^2$ or greater.  We use the GRV LO pion
set \cite{GRVpi} for the pion parton distributions.  We assume $m_s = 500$ MeV.
This constituent quark value modulates $\alpha_s$, keeping
it below unity.  In addition, to avoid going below $Q_0^2$, we assume that
both $\alpha_s$ and the parton distributions are evaluated at scale $\mu
= 2 m_T = 2 \sqrt{p_T^2 + m_s^2}$.  The $x_F$ distribution, obtained by
integrating the differential partonic cross section over the $p_T$ and
rapidity, $y$, of the unobserved quark, selects low $p_T$.

A LO calculation 
provides the basic shape of the $x_F$ distribution.  The shape
does not change
significantly at higher orders, at least to next-to-leading order for
charm and bottom quarks \cite{RV2}.  We will assume that this is
also true for strange quarks.  We further assume that the factorization 
theorem \cite{fact1,fact2,fact3,fact4} 
still holds for perturbative production of strange quarks.
We will address the validity of this assumption when we discuss the model
comparisons to the data.

We prefer to treat the strange quark as heavy rather than as a massless parton
in hard $2 \rightarrow 2$ scatterings, ``jet-like'' 
processes.  There are several
reasons for this.  First, treating the strange quark as a ``jet'' means 
selecting a minimum $p_T$ to keep the cross section finite.  A large minimum
$p_T$ compatible with hard scattering is incompatible with the
assumption of intrinsic production, inherently a low $p_T$
process \cite{VB}.  A jet with a leading 
strange particle can be produced from all $2 \rightarrow 2$ processes in
which a strange quark appears in the final state.  However, 
strange particles can also be produced from the fragmentation of light quark
and gluon jets.  In any case, there is no indication that the strange particles
measured by E791 originate from jets.

The $x_F$ distribution of leading-twist production \cite{VBH2} of
heavy quarks by $gg$ fusion and $q \overline q$ annihilation is 
\be
H = \frac{d\sigma_{\rm lt}^S}{dx_F} & = &
\frac{\sqrt{s}}{2} \int  dz_3\, dy_2\, dp_T^2 
\frac{1}{E_1}\ \frac{D_{S/s}(z_3)}{z_3}\   x_a x_b \bigg( \sum_{q=u,d,s} 
\bigg[ f_q^A(x_a,\mu^2) f_{\overline q}^B(x_b,\mu^2) \nonumber \\
&   & \mbox{} +  f_{\overline q}^A(x_a,\mu^2) f_q^B(x_b,\mu^2) \bigg] 
\frac{d \widehat{\sigma}_{q \overline q}}{d \hat{t}} + f_g^A(x_a,\mu^2) 
f_g^B(x_b,\mu^2) \frac{d \widehat{\sigma}_{gg}}{d \hat{t}} \bigg)
 \, \, ,
\label{ltfus}
\ee
where $a$ and $b$ 
are the projectile and target partons, 1 and 2 are the 
produced strange quarks, and 3 is the final-state strange hadron $S$.
Feynman $x$ is defined as $x_F = 2(m_T/\sqrt{s})\sinh y_3$ where 
$\sqrt{s}$ is the hadron-proton center of mass
energy. The leading order
subprocess cross sections for heavy quark production
can be found in Ref.~\cite{Ellis}.  
The fractional momenta carried by the projectile and target partons, 
$x_a$ and $x_b$,
are $x_a = (m_T/\sqrt{s}) (e^{y_1} + e^{y_2})$ and $x_b = (m_T/\sqrt{s}) 
(e^{-y_1} + e^{-y_2})$ at LO with two massive quarks in the final state.
The strange quark contribution is negligible, less than $0.1$\%.
We equate $d\sigma_{\rm lt}^S/dx_F$ with $H$, the hard scattering
cross section as a useful abbreviation.

The fragmentation functions, $D_{S/s}(z)$, describe the
hadronization of the strange quark into strange hadron $S$.  
Since including the unknown strange
quark fragmentation functions would only add an additional 
degree of uncertainty, we assume that 
\be D_{S/s}(z) = B_S \delta(1-z) \,\, . \label{fusfrag} \ee
A delta function
for fragmentation is in agreement with low $p_T$ charm hadroproduction,
see Ref.~\cite{VBH2}.  If all ten ground state strange hadrons are produced at
the same rate, the normalization, $B_S$, is 0.1.  We may overestimate the
production of some strange hadrons in this way.

\begin{figure}[htpb]
\setlength{\epsfxsize=0.5\textwidth}
\setlength{\epsfysize=0.25\textheight}
\centerline{\epsffile{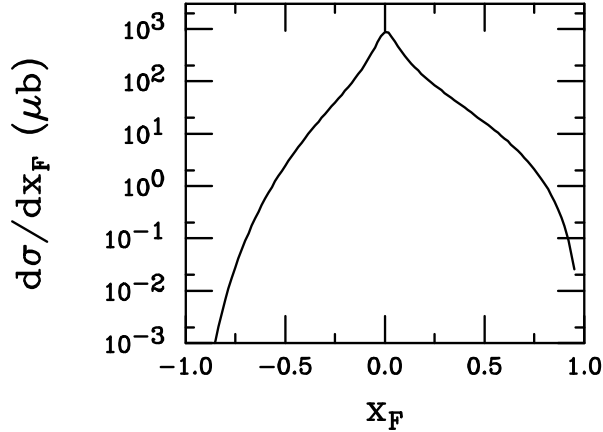}}
\caption[]{Strange quark production by leading-twist fusion in
$\pi^- p$ interactions at 500 GeV.}
\label{fus}
\end{figure}

The LO $x_F$ distribution
for $\pi^- p$ interactions at 500 GeV, the energy of E791
\cite{anjosmex1,anjosmex2}, are shown in
Fig.~\ref{fus}.  No $K$ factors are included.   The forward 
$\pi^- p$ cross section is rather hard, mostly due to the harder pion gluon
distribution.  

Although we have treated the strange quark as massive, we have also checked 
how the $x_F$ distribution would
change if the strange quark was treated as massless and all $2 \rightarrow 2$
scattering channels with an $s$ quark in the final state were included.  
Generally, the additional 
``jet'' production of 
strangeness is through processes
such as
$g s \rightarrow g s$ and $q s \rightarrow qs$ ($\overline q s \rightarrow 
\overline q s$) as well as for the $\overline s$.
Including these ``jet-like'' processes increases the cross section by a factor 
of $4-8$.  While this factor
is not constant, it increases rather slowly with $x_F$ so that the difference
in shape is only important at large $x_F$ where the cross section is decreasing
more rapidly.

\section{Intrinsic Particle Production}
\paragraph{}
 
The wavefunction of a hadron in QCD can be represented as a
superposition of Fock state fluctuations, {\it e.g.}\ $\vert n_v
\rangle$, $\vert n_v q \overline q \rangle$,
$\vert n_v s \overline s \rangle$, $\vert n_v s \overline s q \overline q
\rangle$ \ldots components where $n_v$ are the valence quarks of the hadron.  
The additional $Q \overline Q$ pairs are said to be ``intrinsic''
to the hadron wavefunction. 
When the projectile scatters with the target, the
coherence of the Fock state is broken and the intrinsic fluctuation 
can hadronize.  This hadronization can proceed either by uncorrelated 
fragmentation, as in leading twist
production, or by coalescence with
spectator quarks in the wavefunction \cite{intc1,intc2,BHMT}.  The generic
intrinsic $Q \overline Q$ components are generated by virtual
interactions such as $g g \rightarrow Q \overline Q$ where the
gluons couple to two or more projectile valence quarks. The
Fock states are dominated by configurations with
equal rapidity constituents so that the quarks in an intrinsic state carry 
a larger fraction of the parent hadron momentum \cite{intc1,intc2}.
The momentum boost received by the quarks in an intrinsic state depends on 
their mass and the number of partons in the Fock state probed.

We will calculate strange particle production from $\pi^-(\overline u d) + p
(uud)$ interactions.  At $x_F >0$ the intrinsic fluctuations arise in the pion 
while at $x_F<0$, the proton fragmentation region, the proton can be considered
the source of intrinsic fluctuations.  We assume no strange particle
production by interference between the $\pi^-$ and $p$ Fock states.
 
The probability for an $n$--particle
Fock state is taken to be frame-independent and may be
written as \cite{intc1,intc2}
\be
\frac{dP^n_{\rm iQ}}{dx_i \cdots dx_n} = N_n 
\frac{\delta(1-\sum_{i=1}^n x_i)}{(m_h^2 - \sum_{i=1}^n
(\widehat{m}_i^2/x_i) )^2} \, \, .
\label{icdenom}
\ee
where the subscript ``iQ'' denotes any generic Fock state with an arbitrary 
number, $r$, of intrinsic quark-antiquark pairs.  The pairs could be light,
strange, or heavy.  If the hadron has $n_v$ valence quarks, then the number of
particles in the state is $n = n_v
+ 2r$.  The probability is normalized by $N_n$ where $n=4$ and 5 for 
the minimal $|n_v s \overline s \rangle$ Fock 
configurations in a pion and a proton respectively.  
We consider initial hadron Fock states with up to $r=3$ intrinsic pairs or 
$n = 8$ for mesons and 9 for baryons.
The delta function in the numerator of Eq.~(\ref{icdenom})
conserves longitudinal momentum.  The dominant Fock configurations
are closest to the light-cone energy shell and therefore the invariant
mass, $M^2 = \sum_i \widehat{m}_i^2/ x_i$, is minimized.  The kinematic 
variables of the $i^{\rm th}$ particle in the state are the effective 
transverse mass squared,
$\widehat{m}_i^2 = \langle \vec k^2_{T,i} \rangle 
+ m^2_i$, and $x_i$ is the light-cone momentum
fraction.  Assuming $\langle \vec k_{T, i}^2 \rangle$ is
proportional to the square of the constituent quark mass, we choose
$\widehat{m}_q = 0.45$ GeV and $\widehat{m}_s =
0.71$ GeV \cite{VBH2,VBH1}.

There are two possible ways of producing strange hadrons in the intrinsic
model.  Both require the presence of either a strange valence quark or at
least one intrinsic $s \overline s$ pair in the Fock state configuration.
Strange particles may be produced by uncorrelated fragmentation of a
strange quark in the Fock state, as previously
discussed for leading-twist strangeness production in Section 2.  
No strange hadrons are assumed to be produced by fragmentation of the light
quarks. 
More importantly, the strange quark can hadronize
through coalescence with spectator partons.  If coalescence occurs in the
minimal $|n_v s \overline s \rangle$ Fock state flavor correlations are
introduced between the projectile
and the final-state hadron, giving rise to a leading particle effect.
While coalescence can still occur in higher-$n$ Fock states, the flavor
correlations are weaker and the leading particle effect less important.  In
fact, when $r=3$ both the strange and antistrange hadrons can be produced by
coalescence, giving no leading particle effect.

Uncorrelated fragmentation does not favor the production of one 
strange hadron over any other.  No other valence quarks from the final-state
hadron are necessary to produce a strange hadron by uncorrelated fragmentation.
We assume equal probabilities for all ground
state strange hadrons, independent of their mass and quark content, as in the
treatment of strange particle production at leading twist.  We allow
strange quarks to fragment into strange hadrons and strange antiquarks to
fragment into antistrange hadrons.  We ignore strange particle production by
fragmentation of other light quarks in the configuration.  In this case, if
the strange quark fragments into a kaon, the $K$ distribution is
\be
\frac{d P^{nF}_{\rm iQ}}{dx_{F}} = \int dz \prod_{i=1}^n dx_i
\frac{dP^n_{\rm iQ}}{dx_1 \ldots dx_n} \frac{D_{K/s}(z)}{z} \delta(x_F - z
x_s) \, \, .
\label{icfrag}
\ee
The same distribution is thus valid for all strange particles
produced by uncorrelated fragmentation from a given $n$-particle state.  
The fragmentation function from leading-twist production, 
Eq.~(\ref{fusfrag}), is also used here.   

If the energy denominator is minimized in Eq.~(\ref{icdenom}), as 
required for the intrinsic state to maintain its integrity, fragmentation may
cost more energy than is available to produce the final-state strange 
particle.  We will therefore test the importance of the fragmentation mechanism
in the intrinsic state by comparing our full results to those with $P_{\rm 
iQ}^{nF} = 0$.

For a strange hadron, $S$, to be produced by coalescence, all 
the valence partons of $S$ must be
present in the Fock state.  Since the
multi-particle Fock states are fragile, they can easily coalesce into 
strange hadrons in high-energy, low momentum transfer reactions.   No binding
or mass effects are assumed.  The coalescence contribution to
strange hadron production is then
\be
\frac{d P^{nC}_{\rm iQ}}{dx_F} = \int \prod_{i=1}^n dx_i
\frac{dP^n_{\rm iQ}}{dx_1 \ldots dx_n} \delta(x_F - \sum_{m_v}
x_{S_{m_v}}) \, \,  
\label{iccoalD}
\ee
where $m=2$ for mesons and three for baryons.
The coalescence function is simply a delta function combining the momentum
fractions of the valence quarks of the strange hadron present in the Fock 
state configuration.  It is clear that only a small 
fraction of the strange hadrons can be produced from the minimal 
configuration with $r=1$ ($n=5$ for protons).  
However, coalescence can also occur within Fock state
fluctuations with $r>1$.
Coalescence is calculated the same way in these higher configurations.

Since we wish to study the $x_F$ distributions of all ground state strange and
antistrange hadrons and the asymmetries between them, we include Fock state
configurations with up to $r=3$.  Thus we include all 
possible light quark/strange quark 
combinations in Fock states with $n=9$ for a proton and
with $n=8$ for a pion, allowing coalescence 
production of $\Omega$ and $\overline \Omega$.  
The minimum number of partons needed
to produce a given ground state strange hadron by coalescence is shown in 
Table~\ref{isconfig} along with the required combination of 
$Q \overline Q$ pairs for coalescence production.  

\begin{table}
\begin{center}
\begin{tabular}{|c|c|c|} \hline
 & \multicolumn{2}{c|}{Projectile} \\ 
Final State & $\pi^-(\overline u d)$ & $p (uud)$ \\ \hline
$K^- (\overline u s)$ & $4 (s \overline s)$ & $7 (s \overline s u \overline 
u)$ \\ \hline
$\overline{K^0} (\overline d s)$ & $6 (s \overline s d \overline d)$ 
& $7 (s \overline s d \overline d)$ \\ \hline
$\Lambda (uds)$  & $6(s 
\overline s u \overline u)$& $5(s \overline s)$ \\ \hline
$\Sigma^- (dds)$  & $6 (s \overline s
d \overline d)$ & $7 (s \overline s d \overline d)$ \\ \hline
$\Sigma^+(uus)$ & $8 (s \overline s u \overline u u \overline u)$ & 
$5 (s \overline s)$ \\ \hline
$\Xi^0(uss)$  & $8(s \overline s s \overline s u \overline u)$ & 
$7 (s \overline s s \overline s)$ \\ \hline
$\Xi^-(dss)$  & $6 (s \overline s s \overline s)$
& $7(s \overline s s \overline s)$ \\ \hline
$\Omega(sss)$  & $8(s \overline s s \overline s s \overline s)$
& $9 (s \overline s s \overline s s \overline s)$  \\ \hline \hline
$K^+ (u \overline s)$& $6(s \overline s u \overline u)$ & 
$5(s \overline s)$  \\ \hline
$K^0 (d \overline s)$ & $4 (s
\overline s)$ & $5 (s \overline s)$ \\ \hline 
$\overline \Lambda (\overline u \overline d \overline s)$ 
& $6(u \overline u d \overline d)$ & $9(s \overline s u \overline u d 
\overline d)$  \\ \hline
$\overline{\Sigma^-} (\overline d \overline d \overline s)$ 
& $8 (s \overline s d \overline d d \overline d)$ & $9 (s \overline s d 
\overline d d \overline d)$ \\ \hline
$\overline{\Sigma^+} (\overline u \overline u \overline s)$ 
 & $6(s \overline s u \overline u)$ & $9 (s \overline s u \overline u u 
\overline u)$  \\ \hline
$\overline{\Xi^0}(\overline u \overline s \overline s)$ 
& $6(s \overline s s \overline s)$ & $9 (s \overline s s \overline s u 
\overline u)$  \\ \hline
$\overline{\Xi^-} (\overline d \overline s \overline s)$ 
& $8(s \overline s s \overline s d
\overline d)$ & $9(s \overline s s \overline s d \overline d)$ \\ \hline
$\overline{\Omega} (\overline s \overline s \overline s)$ 
& $8 (s \overline s s \overline s s \overline s)$ & $9 (s \overline s s 
\overline s s \overline s)$ \\ \hline
\end{tabular}
\end{center}
\caption[]{The lowest number of partons in an intrinsic strangeness Fock state
configuration for a strange hadron to be produced by coalescence.}
\label{isconfig}
\end{table}

We now discuss how the probability for the Fock states with $r=1-3$ are
determined.  To remain close to the spirit of the original intrinsic charm
model, we assume that $P_{\rm is}^5$ and
higher Fock state probabilities can be obtained from $P_{\rm ic}^5$ 
by mass scaling, as in Ref.~\cite{GutVogt1}.  
A reanalysis of the EMC charm structure function data with 
next-to-leading order calculations of 
charm electroproduction by both leading-twist photon-gluon fusion
and higher-twist intrinsic charm was shown to be consistent with an
intrinsic charm component in the proton at large
$x_{\rm Bj}$ of $\approx 1$\%\ or less \cite{hsv}.  (See also 
Ref.~\cite{SMT}.)  An earlier analysis
found $P^5_{\rm ic} = 0.31$\% \cite{EMCic1,EMCic2}. 
To be conservative in our estimates
of the intrinsic contribution to strange particle production,
we will always assume that the total probability for a charm quark to arise
from an $|n_v c \overline c \rangle$ state is 0.31\% \cite{hsv,EMCic1,EMCic2},
regardless of the projectile identity, $P_{\rm ic}^5 = P_{\rm 
ic}^4$ for baryons and mesons.  We  
scale $P_{\rm ic}^5$ by the square of the quark transverse masses 
to obtain 
\be 
P^5_{\rm is} & = & \left( \frac{\widehat{m}_c}{\widehat{m}_s} \right)^2 
P^5_{\rm ic} \approx 2\% \label{pis} 
\ee
with $\widehat{m}_c = 1.8$ GeV \cite{VBH1}.  The assumption that 
$P_{\rm ic}^5 = P_{\rm ic}^4$ leads to $P_{\rm is}^5 = P_{\rm is}^4$.

To normalize the probability of states with $r>1$, 
we use the method described in Ref.~\cite{GutVogt1}. 
Data on double charmonium hadroproduction were used to set an upper limit on
the $|n_v c \overline c c \overline c \rangle$ probability: $P_{\rm 
icc}^7 \approx 0.044 \ P_{\rm ic}^5$ \cite{VB2}. Then 
the probabilities for Fock state configurations with $r=2$ ($n=7$ for a proton)
can be fixed.  We begin with \cite{VBlam}
\be
P^7_{\rm icq} \approx \left( \frac{\widehat{m}_c}{\widehat{m}_q}
\right)^2 P^7_{\rm icc} \, \, .
\label{picq}
\ee
Then it follows that
\be
P^7_{\rm isq} & = & \left( \frac{\widehat{m}_c}{\widehat{m}_s}
\right)^2 P^7_{\rm icq}  = 0.704 \, P^5_{\rm is} \, \ , \label{pisq} \\
P^7_{\rm iss} & = & \left( \frac{\widehat{m}_q}{\widehat{m}_s} \right)^2
P^7_{\rm isq} = 0.285 \, P^5_{\rm is} \, \, . \label{piss}
\ee
We take $P_{\rm isu}^7 = P_{\rm isd}^7$.
The relations in Eqs.~(\ref{pisq})-(\ref{piss}) also hold for the $n=6$
pion Fock states. 

There is no guidance from intrinsic charm to normalize the 
Fock configurations with $r=3$ since no triply-charm baryon distributions have
been calculated.  Thus we
assume the same scaling between $P_{\rm issq}^9$ and $P_{\rm iss}^7$ as
between $P_{\rm isq}^7$ and $P_{\rm is}^5$ in Eq.~(\ref{pisq}).  Then
\be
P^9_{\rm issq} = 0.704 \, P^7_{\rm iss} = 0.2 \, P^5_{\rm is} \, \, .
\label{pissq}
\ee
Mass scaling can then be used to obtain the other $r=3$ ($n=9$ for the proton)
probabilities:
\be
P^9_{\rm isss} & = & \left( \frac{\widehat{m}_q}{\widehat{m}_s} \right)^2
P^9_{\rm issq} = 0.081 \, P^5_{\rm is} \, \, , \label{pisss} \\
P^9_{\rm isqq} & = & \left( \frac{\widehat{m}_s}{\widehat{m}_q}
\right)^2 P^9_{\rm issq}  = 0.5 \, P^5_{\rm is} \, \, , \label{pisqq}
\ee
Note that here $P_{\rm issu}^9 = P_{\rm issd}^9$ and $P_{\rm isuu}^9 = P_{\rm
isud}^9 = P_{\rm isdd}^9$.
In this case, we also assume the probabilities of the eight-particle Fock 
configurations are equal to their nine-particle counterparts given in 
Eqs.~(\ref{pissq})-(\ref{pisqq}). 

We have not included the probabilities for intrinsic states with only
light quarks since these do not contribute to strange hadron production in
pion and proton Fock states.  If we consider hyperon beams like the $\Sigma^-$,
these light intrinsic states would also contribute.  The light intrinsic states
must of course be considered when calculating the full probability sum,
\be
P = \sum_{\rm Q} P_{\rm iQ}^5 + \sum_{\rm Q,Q'} P_{\rm iQ Q'}^7
+ \sum_{\rm Q,Q', Q''} P_{\rm iQ Q' Q''}^9 + \cdots \, \,  \label{probsum}
\ee
where ${\rm Q} = u,d,s,\cdots$.
The total probability, $P=1$, sets an upper bound
on $P_{\rm is}^5$ since all other probabilities for lighter and heavier quarks
can be related to it, see Eqs.~(\ref{pisq})-(\ref{pisqq}).
Considering only the light and strange quarks with $r=1-3$, we find
$P \approx 0.4$.  The contributions from heavier quarks do not increase $P$
significantly.  The remainder of the hadron wavefunction would 
include multi-gluon as well as multiquark configurations which we have not
included here.  Thus $P_{\rm is}^5$
cannot be significantly increased to fit data.

The total intrinsic contribution to strange hadron production is a
combination of uncorrelated fragmentation and coalescence.  
We do not consider production from configurations with $r>3$.  
Recall that including still higher Fock states weakens the flavor correlations 
between the strange quarks and the valence quarks.  In fact, only those 
strange hadrons produced
in the Fock configuration with $r=1$, such as the $K^0$, $K^+$, $\Lambda$
and $\Sigma^+$ in the proton, are leading relative
to the remaining strange hadrons.  There will be an asymmetry between 
a $\Xi^-$ in a $|uud s \overline
s s \overline s \rangle$ state and a $\overline{\Xi^-}$ first produced in a 
$|uud s \overline s s \overline s d
\overline d \rangle$ state, albeit not as strong.  Since the 
relative probabilities decrease when 
additional pairs are added to the Fock state, further contributions,
even including coalescence, will have only slightly different 
$x_F$ distributions
than those resulting from uncorrelated fragmentation in a
configuration with lower $n$.  
There is then no longer any advantage in introducing more 
pairs into the configuration because the relative probability will
decrease while the potential gain in momentum is not significant.
However, for coalescence production of the 
$\Omega$,
all possible final-state strange hadrons from states with $r \leq 3$ are 
counted in the total intrinsic probability.  

The unit-normalized probability distributions, 
$(1/P^n_{\rm iQ})(dP^n_{\rm iQ}/dx_F)$, for both uncorrelated fragmentation
and coalescence are given in Appendix A.  
These probability distributions, when properly normalized and weighted, 
will comprise the intrinsic contribution to strange hadron production.  

To calculate the full strange and antistrange hadron, $S$, 
$x_F$ distributions in
the intrinsic model, we include uncorrelated fragmentation of the strange
quark in every state and coalescence from those states with the correct quark
content.  Since the intrinsic probability distributions and the coalescence
mechanism are independent of the final-state mass, the results are identical
for the ground state and higher strange resonances which have the same valence
quarks. We have thus only taken the 10 ground state strange hadrons and 
antihadrons into account.  To conserve uncorrelated fragmentation probability,
we assume that $P_{\rm iQ}^{nF}=0.1$.
For the coalescence contribution, we count the number of 
possible ground state strange and 
antistrange hadron combinations that can be obtained from a given state. 
Each strange hadron or antihadron is assigned a weight, $\xi$, equivalent to
the number of possible ways to produce that hadron from the total number of
$S$ or $\overline S$ hadrons in the state.

In general, the possible number of strange hadrons is greater than
the number of possible antistrange hadrons in 
a given state.  This has the effect
of making, for example, $\overline{\Omega}$ production by coalescence 
more probable than the $\Omega$ in the proton since there are fewer
antistrange hadrons in the final state.  The overall effect is very small
since both the $\Omega$ and $\overline \Omega$ are
only produced from the $|uud s \overline s s \overline s s
\overline s \rangle$ state.  The appropriate $x_F$ distribution from 
coalescence is weighted by the fraction of possible combinations of that
final-state hadron to the total strange hadrons
or antihadrons in each state.  When a strange hadron can be
produced by both fragmentation and coalescence, we take half the sum of the
two contributions to conserve the total probability.  If
$P^{nF} \equiv 0$, only the coalescence weight obtained from 
counting contributes to the total probability.

Finally, to obtain the total probability of each strange hadron in the
intrinsic model, we sum all 
the contributions from all the states.
Thus
\be \frac{dP_S}{dx_F} = \sum_n \sum_{r_u} \sum_{r_d} \sum_{r_s} \beta
\left( \frac{1}{10} \frac{dP_{{\rm i} (r_s {\rm s}) (r_u {\rm u}) (r_d {\rm d})
}^{nF}}{dx_F} + \xi \frac{dP_{{\rm i} (r_s {\rm s}) (r_u {\rm u}) (r_d {\rm d})
}^{nC}}{dx_F} \right) \, \, .
\label{intsum}
\ee
The weight of each state produced by coalescence is
$\xi$ where $\xi = 0$ when $S$ cannot be produced by coalescence from state
$|n_v  r_s(s \overline s) r_u(u \overline u)r_d(d \overline d) \rangle$.  
The parameter $\beta$ is 1 when $\xi = 0$ 
and 0.5 when production by both fragmentation and coalescence is 
possible to conserve probability in each state.  
When we assume coalescence production only, $P^{nF} 
\equiv 0$ and $\beta \equiv 1$.
The number of up, down and strange $Q \overline Q$ pairs 
is indicated by $r_u$, $r_d$ and $r_s$ respectively.  The total, $r_u + r_d
+ r_s = r$, is defined as $r = (n - n_v)/2$ because each $Q$ in an $n$-parton
state is accompanied by a $\overline Q$. For baryon projectiles, $n =5$, 7, 
and 9 while for mesons $n=4$, 6, and 8.  Depending on the value of $n$, $r_i$
can be 0, 1, 2 or 3, {\it e.g.}\ in a $|uud s \overline s d \overline d d 
\overline d \rangle$ state, $r_u = 0$, $r_d = 2$ and $r_s = 1$ with $r = 3$.  
The detailed probability distributions for all strange and 
antistrange hadrons from the
intrinsic states of the $\pi^-$ and $p$ are given in Appendix B.

This method of assigning the probabilities without regard for strange particle
mass is, of course, quite simplistic, 
especially for production by independent fragmentation, but
adequate for testing the general characteristics of the model.  The only way
that baryon number or strangeness number enters the calculation is 
through the choice of $S$.  Other methods
of calculating the relative production rates, such as including the mass in a
statistical fashion, would not distinguish between strange and antistrange
hadrons, as suggested by the data \cite{anjosmex1,anjosmex2}.  
Therefore we make the minimum number of 
assumptions to see if the general framework of the model is correct.

\section{Model Predictions}
 
We now turn to specific predictions of our model for the total strange and
antistrange hadron distributions and the asymmetries between them. 
The $x_F$ distribution for final-state strange hadron $S$ is the sum
of the leading-twist fusion and intrinsic components,
\be
\frac{d\sigma^S_{hN}}{dx_F} = \frac{d\sigma^S_{\rm lt}}{dx_F} +
\frac{d\sigma^S_{\rm iQ}}{dx_F} \, \, .
\label{ismodel}
\ee

The normalization of the production
cross section is determined by
the Fock state probability, the inelastic $hN$ cross section, and a scale 
factor set by the momentum needed to break the coherence of the Fock
state.  
The total intrinsic cross section, $d\sigma^S_{\rm iQ}/dx_F$, 
is related to $dP_S/dx_F$ by
\be
\frac{d\sigma^S_{\rm iQ}}{dx_F} = \sigma_{h N}^{\rm in}
\frac{\mu^2}{4 \widehat{m}_s^2} \frac{dP_S}{dx_F} \, \, .
\label{iscross}
\ee
The scale, $\mu^2$, was fixed at 0.1 GeV$^2$ in intrinsic charm studies
\cite{GutVogt1}. 
Using this scale, $\sigma_{\rm ic} \approx 0.6$ $\mu$b for pions and 0.8
$\mu$b for protons assuming $P_{\rm ic}^5 = 0.3$\%.  This value of $\mu^2$ 
along with $P_{\rm is}^5 = 2$\% gives $\sigma_{\rm is} \approx 0.24$ mb for
pions and 0.3 mb for protons. 
The inelastic $pN$ and $\pi^- N$ cross sections are taken from the Particle
Data Group parameterizations \cite{pdg} and are evaluated at $\sqrt{s'} = 
\sqrt{s}(1 - | x_F |)$ \cite{VBlam}.
Recall that the total probability distributions, $dP_S/dx_F$, for each 
strange hadron $S$ are given in Appendix B.  

To distinguish between the
scenarios with and without fragmentation, we will denote $d \sigma^S_{\rm
iQ}/dx_F$ by $F+C$ for fragmentation and coalescence and $C$ for coalescence
alone.  The total distributions from Eq.~(\ref{ismodel}) are then $H+F+C$ and
$H+C$ respectively where $H$ is the leading-twist cross section in
Eq.~(\ref{ltfus}).  We will also attempt to fit the asymmetry data by scaling
the leading-twist result relative to the intrinsic contribution, denoted by
$H+aC$.  Since we do not know which component should be rescaled, we discuss
the consequences in each case.  We also discuss alternative ways to fit the
data.  Finally, we show the asymmetries without leading-twist fusion or
fragmentation, $C$ only.

\begin{figure}[htpb]
\setlength{\epsfxsize=0.95\textwidth}
\setlength{\epsfysize=0.5\textheight}
\centerline{\epsffile{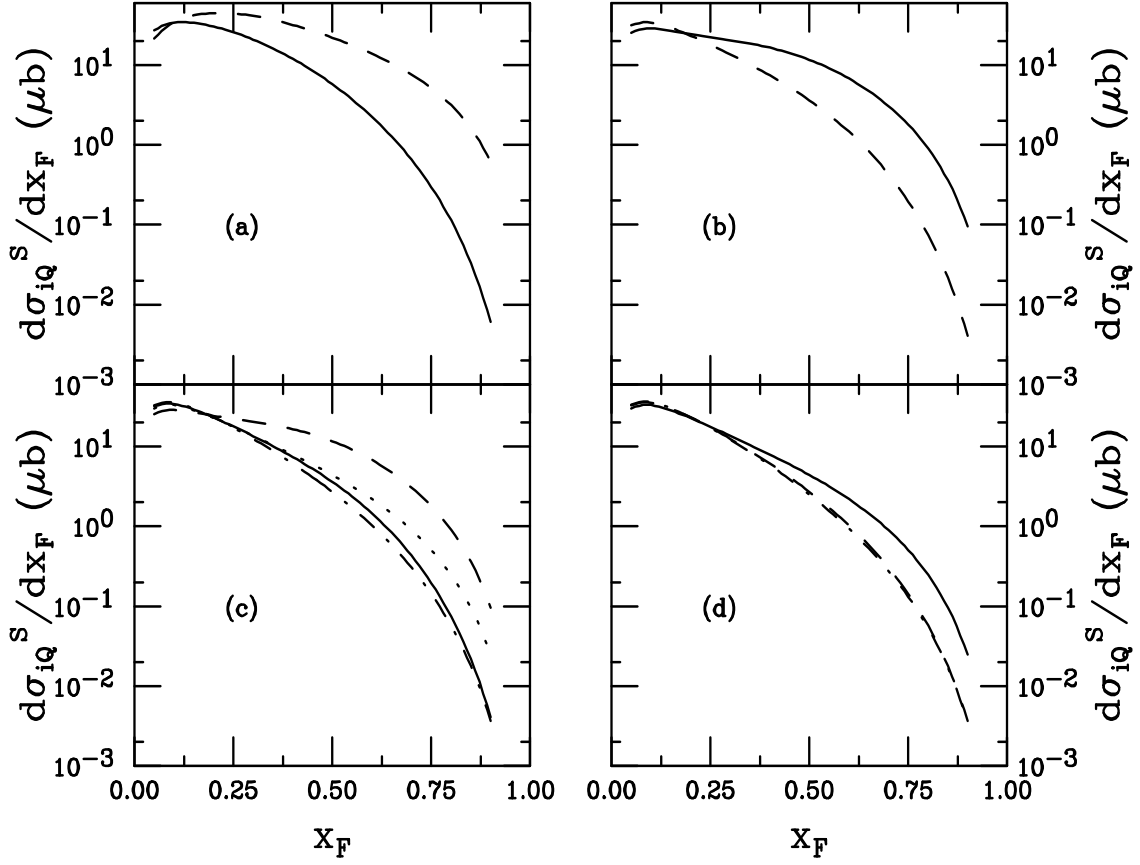}}
\caption[]{The total intrinsic strange and antistrange hadron
production cross sections with a $\pi^-$ projectile.
(a) The solid curve is for $K^+ =
\overline{K^0}$ and the dashed for $K^- = K^0$.  
(b) The $\Lambda = \overline{\Lambda} = \Sigma^-$ (solid) and 
$\overline{\Sigma^-}$ (dashed) distributions
are given.  (c)  The $\Sigma^+$ (solid), $\overline{\Sigma^+}$ (dashed),
$\Xi^0$ (dot-dashed), and $\overline{\Xi^0}$ (dotted) distributions
are shown.  (d) The $\Xi^-$ (solid), $\overline{\Xi^-}$ (dashed) and
$\Omega = \overline{\Omega}$ (dot-dashed) predictions are shown.  }
\label{pitot}
\end{figure}

\begin{figure}[htpb]
\setlength{\epsfxsize=0.95\textwidth}
\setlength{\epsfysize=0.5\textheight}
\centerline{\epsffile{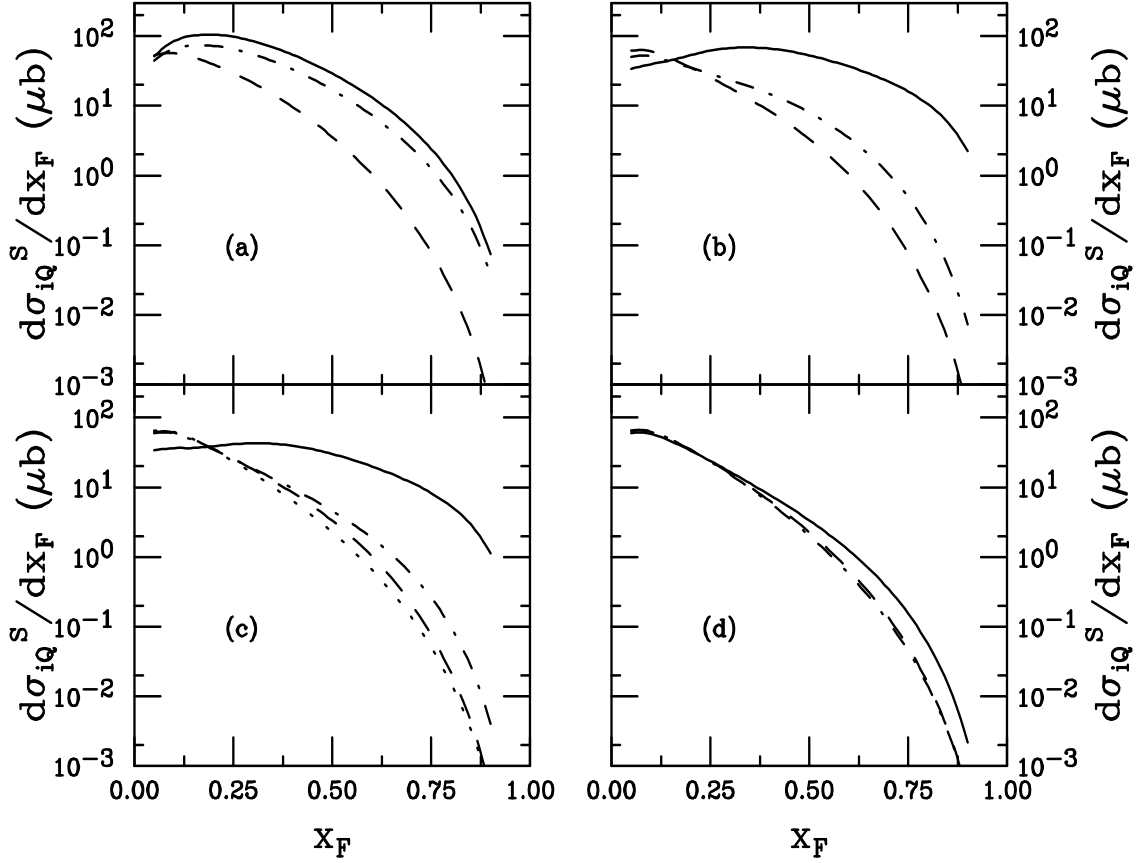}}
\caption[]{The total intrinsic strange and antistrange
hadron
production cross sections with a proton projectile.
(a) The $K^+$ (solid), $K^-$ (dashed), and $K^0$ (dot-dashed) calculations
are presented.  (b) The $\Lambda$ (solid), $\overline{\Lambda}$ and
$\overline{\Sigma^-}$ (dashed), and
$\Sigma^-$ (dot-dashed) distributions
are given.  (c)  The $\Sigma^+$ (solid), $\overline{\Sigma^+}$ (dashed),
$\Xi^0$ (dot-dashed), and $\overline{\Xi^0}$ (dotted) distributions
are shown.  (d) The $\Xi^-$ (solid), $\overline{\Xi^-}$ (dashed), and
$\Omega$ (dot-dashed) predictions are shown.  The $\overline{\Omega}$
distribution is indistinguishable from the $\Omega$ distribution here
even though the two are not identical.}
\label{ptot}
\end{figure}

\begin{figure}[htpb]
\setlength{\epsfxsize=0.95\textwidth}
\setlength{\epsfysize=0.5\textheight}
\centerline{\epsffile{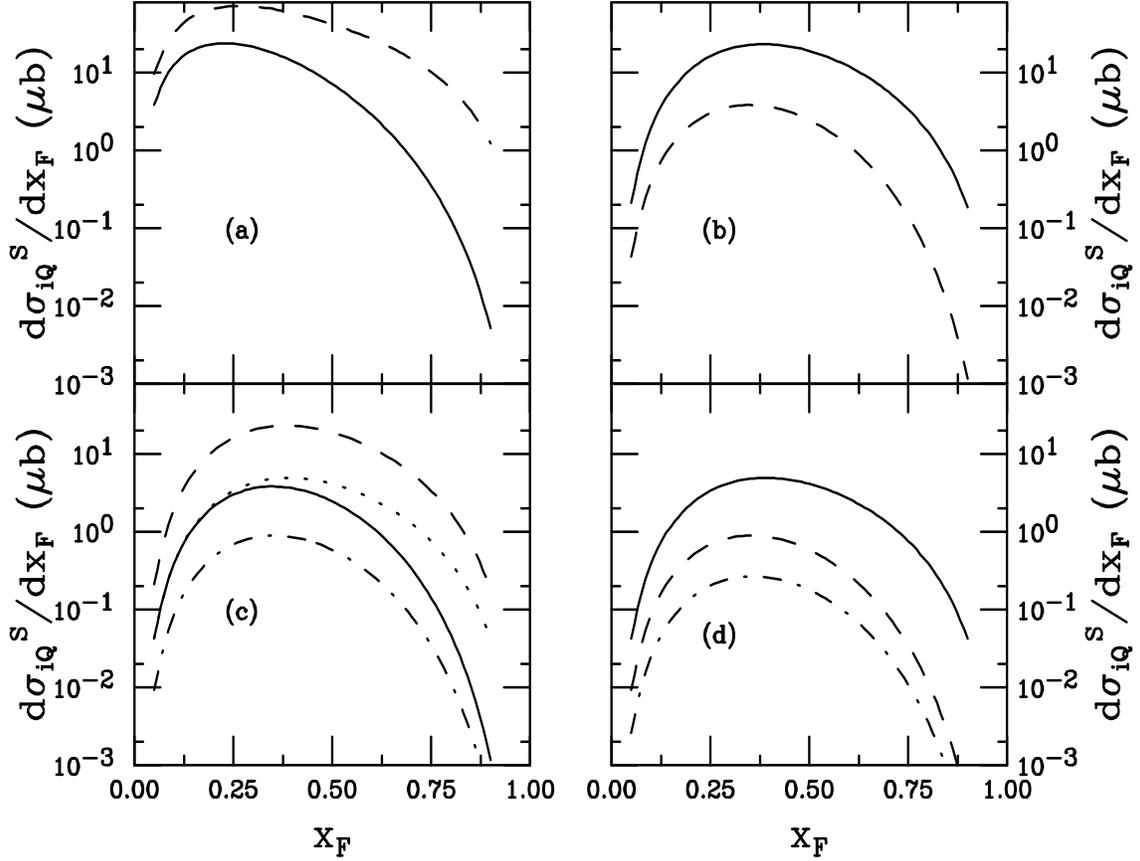}}
\caption[]{The total intrinsic strange and antistrange
hadron
production cross sections with a $\pi^-$ projectile
including coalescence only.
(a) The solid curve is for $K^+ =
\overline{K^0}$ and the dashed for $K^- = K^0$.  
(b) The $\Lambda = \overline{\Lambda} = \Sigma^-$ (solid) and 
$\overline{\Sigma^-}$ (dashed) distributions
are given.  (c)  The $\Sigma^+$ (solid), $\overline{\Sigma^+}$ (dashed),
$\Xi^0$ (dot-dashed), and $\overline{\Xi^0}$ (dotted) distributions
are shown.  (d) The $\Xi^-$ (solid), $\overline{\Xi^-}$ (dashed) and
$\Omega = \overline{\Omega}$ (dot-dashed) predictions are shown.  }
\label{pitotco}
\end{figure}

\begin{figure}[htpb]
\setlength{\epsfxsize=0.95\textwidth}
\setlength{\epsfysize=0.5\textheight}
\centerline{\epsffile{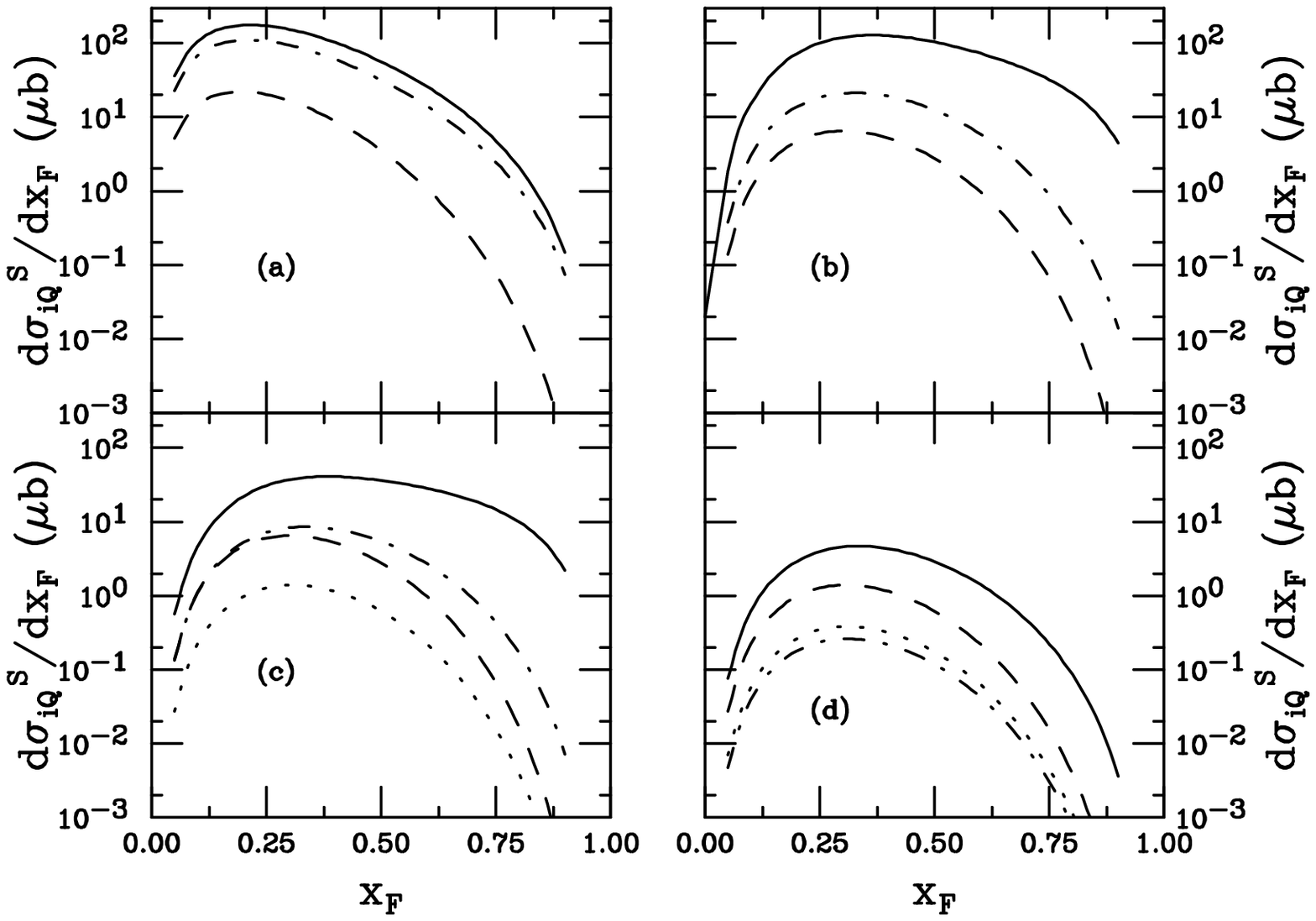}}
\caption[]{The total intrinsic strange and antistrange
hadron
production cross sections with a proton projectile
including coalescence only.
(a) The $K^+$ (solid), $K^-$ (dashed), and $K^0$ (dot-dashed) calculations
are presented.  (b) The $\Lambda$ (solid), $\overline{\Lambda}$ and
$\overline{\Sigma^-}$ (dashed), and
$\Sigma^-$ (dot-dashed) distributions
are given.  (c)  The $\Sigma^+$ (solid), $\overline{\Sigma^+}$ (dashed),
$\Xi^0$ (dot-dashed), and $\overline{\Xi^0}$ (dotted) distributions
are shown.  (d) The $\Xi^-$ (solid), $\overline{\Xi^-}$ (dashed),
$\Omega$ (dot-dashed), and $\overline \Omega$ (dotted) predictions are 
shown.}
\label{ptotco}
\end{figure}

We give $d\sigma_{\rm iQ}^S/dx_F = F+C$
from $\pi^- p$ interactions at 500 GeV in Figs.~\ref{pitot} and \ref{ptot}.  
The results
in Fig.~\ref{pitot} correspond to $x_F >0$
while those in Fig.~\ref{ptot} correspond to $x_F<0$.  The distributions 
$d\sigma_{\rm iQ}^S/dx_F = C$ 
are given in Figs.~\ref{pitotco} and \ref{ptotco}.  
The energy dependence enters only through $\sigma_{hN}^{\rm in}$ which
sets the relative normalization at $x_F \sim 0$.
Some of the intrinsic distributions are equal for a given projectile.
The largest number
of distributions are related for the $\pi^-$ since the pion has both a valence
quark and a valence antiquark.  Then, 
\begin{eqnarray} 
\frac{d\sigma^{K^-}_{\rm iQ}}{dx_F} & = & \frac{d\sigma^{K^0}_{\rm 
iQ}}{dx_F} \,\,\,\,\,\,\, 
\,\,\,\,\,\,\, \,\,\,\,\,\,\, \,\,\,\,\,\,\, 
\,\,\,\,\,\,\, \,\,\,\,\,\,\, \,\,\,\,\,\,\, 
\,\,\,\,\,\,\, \,\,\,\,\,\,\, \,\,\,\,\,\,\, 
\frac{d\sigma^{K^+}_{\rm iQ}}{dx_F} \,\, = \,\, 
\frac{d\sigma^{\overline{K^0}}_{\rm iQ}}{dx_F}  \nonumber \\
\frac{d\sigma^{\Lambda}_{\rm iQ}}{dx_F} & = & 
\frac{d\sigma^{\overline{\Lambda}}_{\rm iQ}}{dx_F}
\,\, = \,\, \frac{d\sigma^{\Sigma^-}_{\rm iQ}}{dx_F} \,\, = \,\,
\frac{d\sigma^{\overline{\Sigma^+}}_{\rm iQ}}{dx_F}  
\,\,\,\,\,\,\,  \,\,\,\,\,\,\,
 \,\,\,\,\,\,\,
\frac{d\sigma^{\Sigma^+}_{\rm iQ}}{dx_F} \,\, = \,\, 
\frac{d\sigma^{\overline{\Sigma^-}}_{\rm iQ}}{dx_F} \label{pionequiv} \\
\frac{d\sigma^{\overline{\Xi^-}}_{\rm iQ}}{dx_F} & = & 
\frac{d\sigma^{\Xi^0}_{\rm iQ}}{dx_F}
 \,\,\,\,\,\,\, \,\,\,\,\,\,\, \,\,\,\,\,\,\,  \,\,\,\,\,\,\, 
\,\,\,\,\,\,\, \,\,\,\,\,\,\, \,\,\,\,\,\,\, 
\,\,\,\,\,\,\, \,\,\,\,\,\,\, \,\,\,\,\,\,\, \,
\frac{d\sigma^{\Xi^-}_{\rm iQ}}{dx_F} \,\, = \,\,
\frac{d\sigma^{\overline{\Xi^0}}_{\rm iQ}}{dx_F} \, \, . \nonumber
\end{eqnarray} 
The relations in the left column of Eq.~(\ref{pionequiv}) are for
``leading''
particles with valence
quarks in common with the projectile while those in the right column are
``nonleading''.  In addition, $d\sigma^\Omega_{\rm iQ}/dx_F = 
d\sigma^{\overline \Omega}_{\rm iQ}/dx_F$.
The antistrange hadron distributions from baryon projectiles
are more likely than the strange hadron distributions to be the same
since only strange hadrons can share valence quarks with the projectile.  
Thus, for protons we find that 
\begin{eqnarray}
\frac{d\sigma^{K^-}_{\rm iQ}}{dx_F} & = & 
\frac{d\sigma^{\overline{K^0}}_{\rm iQ}}{dx_F} 
\nonumber \\ 
\frac{d\sigma^{\overline{\Sigma^-}}_{\rm iQ}}{dx_F} & = & 
\frac{d\sigma^{\overline{\Sigma^+}}_{\rm iQ}}{dx_F} \,\, = \,\,
\frac{d\sigma^{\overline{\Lambda}}_{\rm iQ}}{dx_F} \label{proequiv} \\ 
\frac{d\sigma^{\overline{\Xi^-}}_{\rm iQ}}{dx_F} & = & 
\frac{d\sigma^{\overline{\Xi^0}}_{\rm iQ}}{dx_F} \, \, .  \nonumber
\end{eqnarray}
All the equalities in Eq.~(\ref{proequiv}) are for ``nonleading''
particles.

The $K^- = K^0$ distributions in Fig.~\ref{pitot}(a) are the hardest strange
hadron distributions from the $\pi^-$, as expected from Table~\ref{avexfpi}.
The $\Lambda \, (\Sigma^0)$, $\overline \Lambda$, $\Sigma^-$ and 
$\overline{\Sigma^+}$ are the hardest strange baryon distributions, followed by
the $\overline{\Xi^0}$ and $\Xi^-$.
The pion-induced strange hadron distributions in Fig.~\ref{pitot} are all 
relatively harder than those from the proton, shown in Fig.~\ref{ptot}.  
Due to the pion valence 
antiquark, the antistrange hadron distributions can sometimes be harder than 
the strange hadron distributions, compare the $\overline{\Sigma^+}$ and the
$\Sigma^+$ distributions as well as the $\overline{\Xi^0}$ and the $\Xi^0$
distributions in Fig.~\ref{pitot}(c).  

The $\Lambda$ in Fig.~\ref{ptot}(b) and the $\Sigma^+$ in 
Fig.~\ref{ptot}(c) have the hardest strange baryon distributions in a proton
projectile.  The distributions are relatively flat because the final-state
hadrons both share two valence quarks, $ud$ and $uu$ respectively, 
with the proton. Both can be produced from Fock states with $n=5$.  
The $\Lambda$ distribution is the hardest of the two since
either of the two $u$ valence quarks in the proton can be paired with the
$d$ in the minimal state while the pairing of the two $u$ valence quarks
can only happen once.  The counting differences also occur in the higher Fock
configurations, but then it is equally likely that the strange quark is paired
with a valence or a sea quark since the model makes no distinction between 
their distributions.  When only a single proton valence quark is 
shared with the final-state hadron, the average $x_F$ in the intrinsic model 
is much lower, as can be seen in a comparison between the $\Lambda$ and 
$\Sigma^-$ distributions in Fig.~\ref{ptot}(b).  A comparison of the 
$\Xi^0$ distribution in Fig.~\ref{ptot}(c) and the $\Xi^-$ distribution in 
Fig.~\ref{ptot}(d) shows that the combinatoric effect of two valence $u$ quarks
against a single valence $d$ quark also affects the doubly strange baryons.
The $\Xi^0$ distribution is harder than the $\Xi^-$ even though neither can be
produced in the minimal Fock configuration. 
Likewise, the $K^+$ and $K^0$ share a single valence quark with
the proton.  The combinatorial effect also holds for the mesons so that the
$K^+$ has a harder distribution than the $K^0$, as
seen in Fig.~\ref{ptot}(a).   The antistrange meson and baryon distributions 
are all similar.

The corresponding results with coalescence alone are shown in
Figs.~\ref{pitotco} and \ref{ptotco}.  Now the low to intermediate 
$x_F$ fragmentation contributions are missing, resulting in much lower
intrinsic contributions in that region.  The relative normalization
at high $x_F$ also changes since including 
fragmentation reduces the coalescence contribution by a factor of two
in Eq.~(\ref{intsum}) since
$\beta = 0.5$ with fragmentation and $\beta = 1$ with coalescence alone.
This can be seen, for example, in the $\Lambda$ distribution in 
Fig.~\ref{pitotco}(b).  On the other hand, leaving out
fragmentation from states with $r=1$ causes the 
$x_F$ distributions of strange hadrons produced by coalescence only
in Fock states with $r=3$,
such as the $\overline{\Sigma^+}$, to decrease more rapidly at high $x_F$
because the tail of the fragmentation distribution from the $r=1$ 
state gives a larger contribution at high $x_F$ 
than coalescence from the $| \overline u d
s \overline s u \overline u u \overline u \rangle$ state.  These results are
typical for all strange antibaryon distributions 
where coalescence can occur in only a
single Fock state.  In these cases, the intrinsic model $x_F$ distribution
is simply the corresponding distribution from 
Figs.~\ref{probpi46}-\ref{probp9} normalized to the cross section as in
Eq.~(\ref{iscross}).  Here the differences in the distributions due to the
weight factors $\xi$ in Eq.~(\ref{intsum}) are clearly visible, as seen in
the separation between the $\Omega$ and $\overline \Omega$ distributions in
Fig.~\ref{ptotco}(d).  

Our complete results are the sum of the distributions shown in 
Figs.~\ref{pitot}-\ref{ptotco} with the 
leading-twist production in Fig.~\ref{fus}.  A
comparison of these figures shows the general trends we can expect for 
$d\sigma^S_{hN}/dx_F$ in Eq.~(\ref{ismodel}).  
The intrinsic cross sections in Fig.~\ref{pitot} are 
about 100 times smaller than the leading
twist cross section at $x_F \sim 0$ and would
only become important for hadrons sharing valence 
quarks with the pion unless considerable rescaling is needed to fit the data.  

It is often difficult to obtain high statistics on single hadron distributions,
especially at large $x_F$.  Therefore, a more 
quantitative way to study very similar strange hadron distributions
is through the asymmetry, defined as
\be
A_S(x_F) = \frac{d\sigma^S/dx_F - d\sigma^{\overline S}/dx_F}{d\sigma^S/dx_F +
d\sigma^{\overline S}/dx_F} \, \, 
\label{asymdef}
\ee
where again $S$ represents a strange hadron and $\overline S$ its antistrange 
counterpart.  Note that
we choose to form the asymmetry between strange and antistrange hadrons rather
than defining ``leading'' and ``nonleading'' particles
for each projectile because the
definition of ``leading'' 
may change from one projectile to another.  For example, the  
$K^+(u \overline s)$ is leading in the proton but not in the $\pi^-$ since
the $\pi^-$ has no valence $u$ quark.  

The asymmetries $A_{\Lambda}$, $A_{\Xi^-}$, and $A_{\Omega}$ 
have been measured in $\pi^-$-induced 
interactions at 500 GeV \cite{anjosmex1,anjosmex2}.  
The measurements are around $|x_F|<0.1$.  In 
the forward direction, $x_F >0$, $A_\Lambda$ 
and $A_\Omega$ are independent
of $x_F$ with $A_\Lambda \approx A_\Omega \approx 0.1$ while $A_{\Xi^-}$ 
increases with $x_F$
to $A_{\Xi^-} \approx 0.2$.
On the other hand, at negative $x_F$, only $A_\Omega$ is independent
of $x_F$.  The other asymmetries increase as $x_F$ decreases, approaching 
$A_\Lambda \approx 0.4$ and $A_{\Xi^-} \approx 0.3$ at $x_F = - 0.1$.
The data are inconsistent with PYTHIA which produces essentially no asymmetry
between the particle/antiparticle combinations at forward $x_F$ while at
negative $x_F$, only $A_\Lambda$ is increasing significantly
although less rapidly than the data \cite{anjosmex1,anjosmex2}.  
At negative $x_F$, $A_{\Xi^-}$
remains small while $A_\Omega$ becomes negative.
Note that even if PYTHIA was tuned to reproduce the asymmetries 
at $x_F \approx 0$, the behavior would remain inconsistent with the
data.  On the other hand, the trends of the data are consistent with
recombination models which predict $A_\Lambda > A_{\Xi^-} 
> A_\Omega$ at negative $x_F$ and $A_{\Xi^-} 
> A_\Lambda \sim A_\Omega$ at forward $x_F$ \cite{anjosmex1,anjosmex2}.
The coalescence contributions to the intrinsic model have the same general 
trends as the recombination model.  However, the distributions may differ in
detail.

We can compare
our model calculations with the E791 results.  In the
case of a $\pi^-$ projectile, positive $x_F$ is the beam fragmentation region
and, in the intrinsic model, the strange hadrons are intrinsic to the pion.
Negative $x_F$ corresponds to the target fragmentation region which is modeled
as intrinsic production from a proton.
Thus to form the asymmetry at negative $x_F$, we take the 
proton-induced intrinsic
probability distributions and sum these with the $\pi^- p$ leading-twist
calculation.  We therefore give results in each $x_F$ region separately.

An asymmetry of $\sim 0.14$ at $x_F \sim 0$, as seen by E791 
\cite{anjosmex1,anjosmex2}, suggests that the
strange baryon production cross sections are about 30\% larger than those of 
antistrange baryons.  This initial asymmetry between baryons and antibaryons
at leading twist could arise from associated production of strange baryons with
antistrange kaons, for example 
$\pi^- p \rightarrow \Sigma^+ K^0 \pi^-$, $\Lambda K^+ 
\pi^-$, or $\Sigma^- K^+ \pi^0$.  Thus strange baryon production only 
requires that one or more kaons be 
produced to conserve strangeness and baryon number.  
However, when an antistrange baryon is
produced, both strangeness and baryon number conservation require at least 
two baryons\footnote{The baryons
need not be strange if kaons are also produced.} to be produced with it, for 
example $\pi^- p \rightarrow \Lambda \overline \Lambda n$.  These additional
baryons increase the kinetic energy threshold by 3.5 GeV \cite{cap}.  
The beam energy may not be high enough for the 
increased energy threshold of antistrange baryon production to be neglected.
In this situation, it is easy to imagine a 30\% or greater strange baryon 
and/or antistrange kaon 
enhancement which manifests itself as a nonzero asymmetry at $x_F \sim 0$.

To check how well our model can reproduce the
trends of the data without tuning, 
we have assumed that the leading twist fusion cross 
section is 30\%
larger for all strange relative to antistrange baryons.  We also assume that,
because $K^+(u \overline s)$ and $K^0(d \overline s)$ production is favored by
associated production over $K^-(\overline u s)$ and $\overline{K^0} (\overline
d s)$, the $K^+$ and $K^0$ cross sections are also 30\% larger than the $K^-$
and $\overline{K^0}$.  This is more reasonable than forcing exact strangeness
conservation in the model because the $\pi^- p \rightarrow K^0 \overline{K^0}
n$ kinetic threshold is only 360 MeV greater than that of the 
$\Lambda K^+ \pi^-$ final
state and $K \overline K$ pair production would moderate the $K^0$ over
$\overline{K^0}$ enhancement from strangeness conservation.
There is then no 
exact strangeness conservation in our perturbative 
leading-twist calculation.  Since a model of the exclusive strangeness 
production channels is inherently nonperturbative, it is beyond the scope
of our leading-twist calculation.  Therefore an assumption of an overall
asymmetry of 30\% for strange baryons and antistrange mesons
is more reasonable than assuming
exact strangeness conservation without a complete knowledge of
the associated production channels.  Therefore the real
$A_S$ may change 10-20\% at
$x_F \approx 0$ with energy and final-state particle.

On the other hand, exact strangeness conservation is
required in the intrinsic model since $s$ and $\overline s$ quarks must be
added to the Fock state in pairs.  Baryon
production by coalescence is naturally favored over antibaryon production in
the intrinsic model, as seen by inspection of 
Eqs.~(\ref{piprobkp})-(\ref{pprobomb}).  No other initial asymmetry need be
considered.
Therefore, we assume an initial asymmetry only in the leading-twist 
calculation. 

\begin{figure}[htpb]
\setlength{\epsfxsize=0.95\textwidth}
\setlength{\epsfysize=0.5\textheight}
\centerline{\epsffile{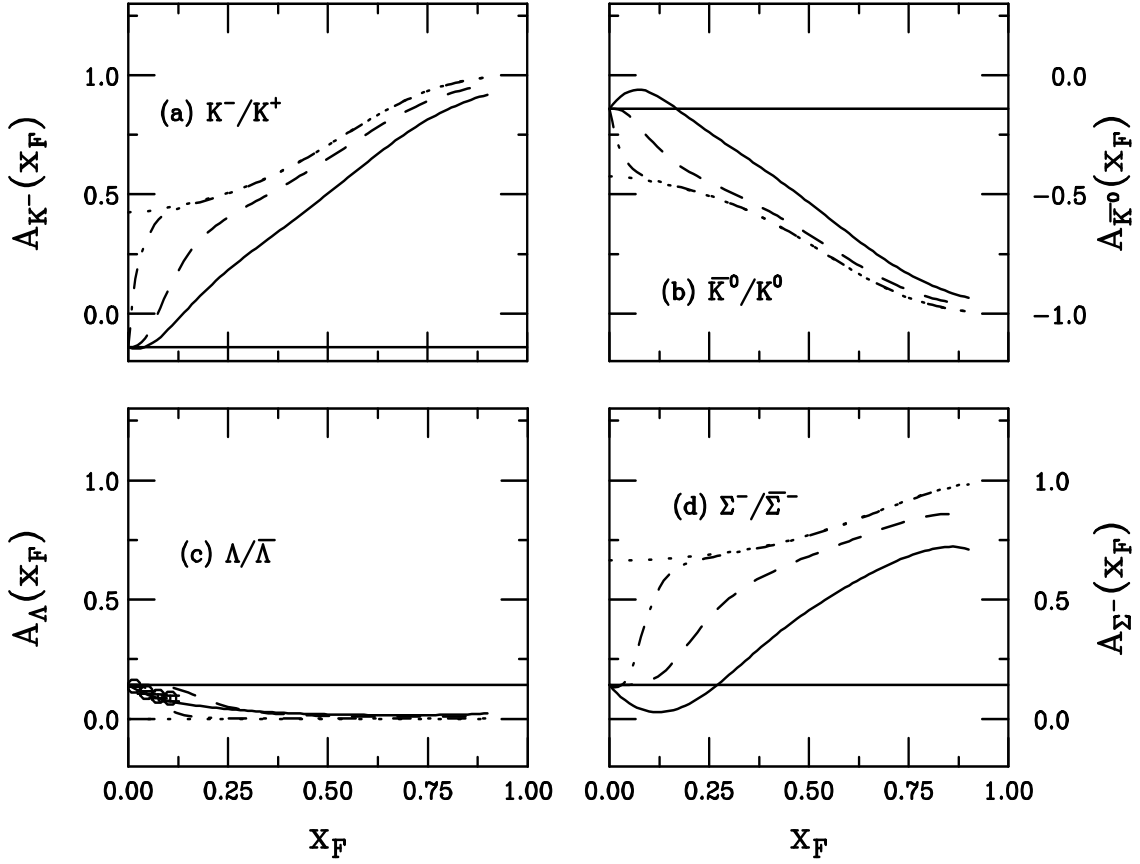}}
\caption[]{ Model asymmetries for $\pi^- p$ interactions at 500 GeV with
$H+F+C$ (solid), $H+C$ (dashed), $H+aC$ with $a=40$
(dot-dashed) and $C$ alone (dotted) are shown 
for (a) $A_{K^-}$, (b)
$A_{\overline{K^0}}$, (c) $A_{\Lambda}$ and (d) 
$A_{\Sigma^-}$. The horizontal solid curve is the asymmetry from leading-twist
production alone.  The E791 data \cite{anjosmex1,anjosmex2} 
on $A_\Lambda$ are also shown. }
\label{api1}
\end{figure}

\begin{figure}[htpb]
\setlength{\epsfxsize=0.95\textwidth}
\setlength{\epsfysize=0.5\textheight}
\centerline{\epsffile{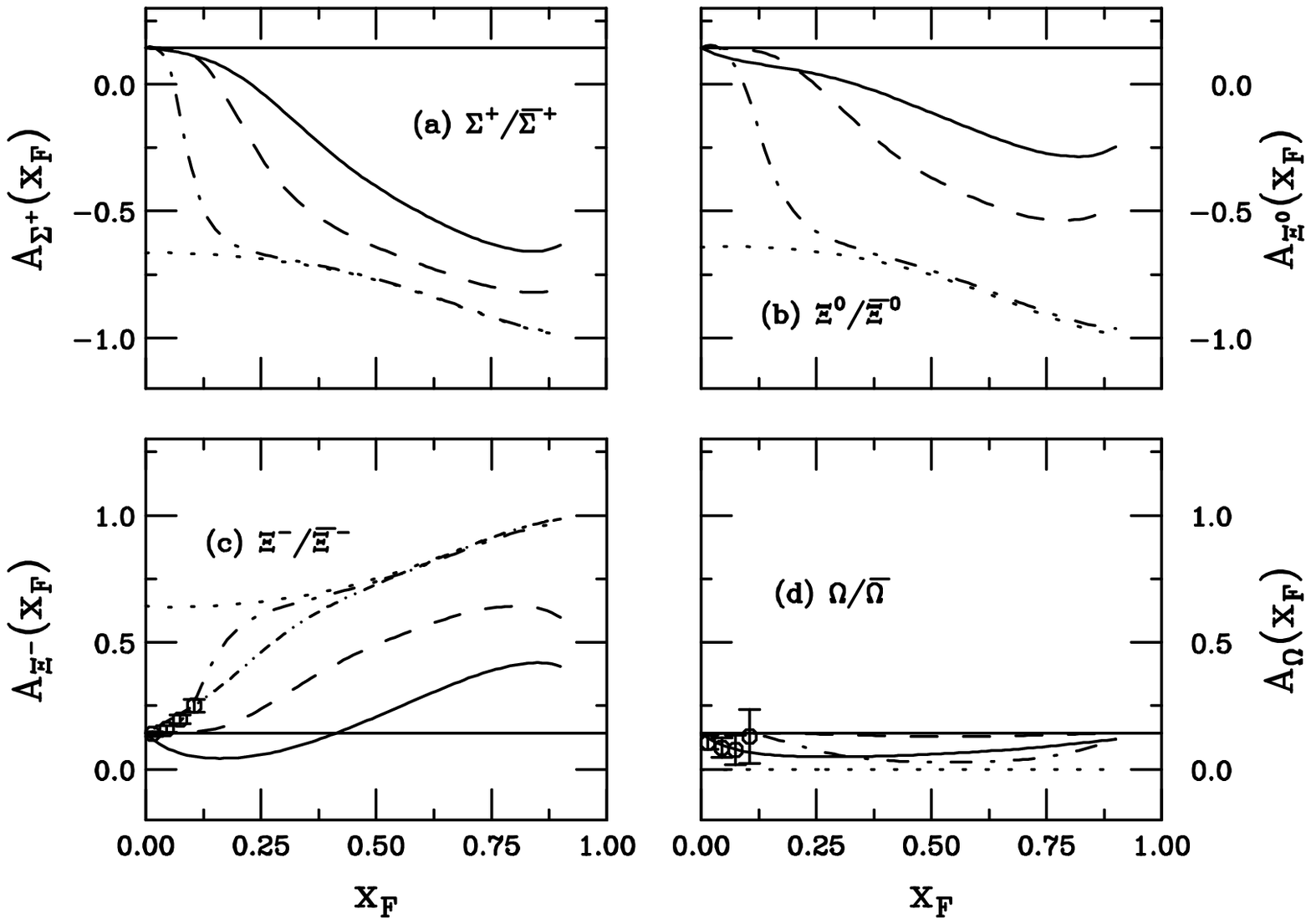}}
\caption[]{ Model asymmetries for $\pi^- p$ interactions at 500 GeV with 
$H+F+C$ (solid), $H+C$ (dashed), $H+aC$ with $a=40$
(dot-dashed) and $C$ alone (dotted) are shown 
for (a) $A_{\Sigma^+}$, (b) 
$A_{\Xi^0}$, (c) $A_{\Xi^-}$ and (d) 
$A_\Omega$.  The result with the modified $\overline{\Xi^-}$
distributions is shown in the dot-dot-dash-dashed curve in (c).
The horizontal solid curve 
is the asymmetry from leading-twist
production alone.   The E791 data \cite{anjosmex1,anjosmex2} 
on $A_{\Xi^-}$ and $A_\Omega$
are also shown.}
\label{api2}
\end{figure}

We first present the asymmetries for $\pi^- p$ interactions at 500 GeV in
Figs.~\ref{api1} and \ref{api2}. In the pion fragmentation region,
$A_{\Sigma^+}$ and $A_{\Xi^0}$ are
negative at large $x_F$ due to the $\overline u$ valence quark in the $\pi^-$.
The asymmetry $A_{\Sigma^+}$ is larger than $A_{\Xi^0}$ due to both the lower 
probabilities of the higher Fock states and the reduced average $x_F$ for a
doubly strange hadron.  On the other hand, $A_{\Sigma^-}$ and $A_{\Xi^-}$
are positive, reflecting the $d$ valence quark of the $\pi^-$.  The situation
is reversed for the meson asymmetries, $A_{K^-}$, associated with the
$\overline u$ valence quark, is positive while $A_{\overline{K^0}}$, associated
with the $d$ valence quark, is negative.  This is simply because the
asymmetries are defined as the difference between hadrons with $s$ and
$\overline s$ quarks.  Since the `leading' particles would be the
$K^-(\overline u s)$ and $K^0(d \overline s)$, with $s$ and $\overline s$ 
quarks respectively, $A_{K^-}$ is positive and $A_{\overline{K^0}}$ is
negative.  The meson 
asymmetries are larger than the baryon asymmetries because the $K^-$ and $K^0$
can be produced by coalescence already in the minimum Fock state configuration.

Observe that $A_{\Lambda}$ and $A_\Omega$ are virtually flat 
and would be exactly zero
if we had assumed either exact strangeness conservation for both
the leading twist and the intrinsic calculations or an initial asymmetry 
between strange and antistrange hadrons in both models.  The fact that the
asymmetry decreases with $x_F$ is because the 
assumed 30\% difference between the particle and antiparticle cross sections
in the leading-twist calculation becomes less important as $x_F$ increases
and the intrinsic contribution begins to dominate.

We also note 
that even though, for example in Figs.~\ref{pitot} and \ref{pitotco}, 
the intrinsic $K^+$ and $\overline{K^0}$
$x_F$ distributions are equal and the $K^-$ and $K^0$ $x_F$ distributions are
equal, $|A_{K^-}|$ and $|A_{\overline{K^0}}|$ are not equal.  This 
is because the particle/antiparticle enhancement factor of 
1.3 is applied to the
$K^+$ and $K^0$, resulting in different asymmetries at low $x_F$.  We find
the same differences at low $x_F$ between $A_{\Sigma^-}$ and $A_{\Sigma^+}$
as well as between $A_{\Xi^0}$ and $A_{\Xi^-}$.

We now compare the results with $H+F+C$ and $H+C$.
As noted earlier, when the intrinsic state is assumed
to undergo independent fragmentation as at leading-twist, the probability
for the final state strange hadron to be produced within any Fock 
state is evenly divided between
fragmentation and coalescence to conserve probability, hence $\beta = 0.5$ in 
Eq.~(\ref{intsum}).  This division results in a ``dip''
in the asymmetry at low $x_F$, as seen for example in $A_{\Sigma^-}$ in 
Fig.~\ref{api1}(d).  (Similar results were observed for charm hadron
asymmetries,
see Ref.~\cite{VB}.)  Thus, turning off intrinsic fragmentation has the general
result of increasing the asymmetry more rapidly at low $x_F$.  
The effect is largest for
$A_{K^-}$ and $A_{\overline{K^0}}$ since the $K^-$ and $K^0$ are both produced
in the lowest state with $n=4$.  
For particles produced by coalescence in higher Fock states,
the increase is slower, such as for $A_{\Xi^-}$ in Fig.~\ref{api2}(c).  
Since
the $\Xi^-$ is first produced by coalescence in the $|\overline u d s \overline
s s \overline s \rangle$ state, the asymmetry does not begin to increase
until $x_F > 0.1$.  This rather slow turn on 
is not in accord with the E791 data
but is in better agreement with the data than the calculations including
intrinsic-model fragmentation.  The $A_\Lambda$ and $A_\Omega$ results do not
change significantly if fragmentation is neglected in the intrinsic model.
Thus, the results in Figs.~\ref{api1} and \ref{api2} reflect the general trend
of the E791 data but $A_{\Xi^-}$ does not increase as 
rapidly as the data, even with intrinsic-model fragmentation turned off.

Regardless of whether or not fragmentation is included does not significantly
affect the agreement of our calculations with the data for $A_\Lambda$ and
$A_{\Omega}$.  However, it is clear that the ``dip'' in $A_{\Xi^-}$ caused by
fragmentation goes in the opposite direction from the data.  Even leaving
fragmentation out does not cause $A_{\Xi^-}$ to increase at low $x_F$.  Thus to
find agreement with this data, we rescale the intrinsic contribution, $H+aC$.
We could equally well rescale the fusion while leaving the intrinsic fixed to
achieve the same effect.  Choosing $a=40$ gives the agreement shown in the
dot-dashed curves.  Comparing this result to that with the asymmetry in the
intrinsic model with $C$ only shows that by $x_F \sim 0.4$, the $H+aC$ result
behaves like that with $C$ alone.  At this point, the relative intrinsic
contribution is large enough to produce a second peak in the $x_F$ distribution
of $H+aC$, as seen in PYTHIA calculations of charm hadrons, see
Ref.~\cite{GutVogt1}.  This convergence of the two results appears at lower
$x_F$ for singly strange baryons and $K$ mesons, $x_F \sim 0.25$ and 0.10
respectively.  Only in these more extreme cases does $A_S \rightarrow 1$ for
most strange hadrons.

\begin{table}
\begin{center}
\begin{tabular}{|c|c|c|c|c|} \hline
Final State & $H+F+C$ & $H+C$ & $H+aC$ & $C$ only \\ \hline
$K^- (\overline u s)$ 
& 0.259 & 0.307 & 0.357 & 0.359 \\ \hline
$\overline{K^0} (\overline d s)$ 
& 0.178 & 0.201 & 0.286 & 0.291 \\ \hline
$\Lambda (uds)$ 
& 0.200 & 0.253 & 0.411 & 0.420 \\ \hline
$\Sigma^- (dds)$
& 0.200 & 0.253 & 0.411 & 0.420 \\ \hline
$\Sigma^+(uus)$ 
& 0.152 & 0.144 & 0.332 & 0.378 \\ \hline
$\Xi^0(uss)$ 
& 0.147 & 0.124 & 0.256 & 0.381 \\ \hline
$\Xi^-(dss)$ 
& 0.158 & 0.163 & 0.386 & 0.425 \\ \hline
$\Omega(sss)$ 
& 0.145 & 0.119 & 0.184 & 0.384 \\ \hline \hline
$K^+ (u \overline s)$  
& 0.170 & 0.189 & 0.285 & 0.291 \\ \hline
$K^0 (d \overline s)$ 
& 0.246 & 0.292 & 0.357 & 0.359 \\ \hline
$\overline \Lambda (\overline u \overline d \overline s)$ 
& 0.211 & 0.275 & 0.414 & 0.420 \\ \hline
$\overline{\Sigma^-} (\overline d \overline d \overline s)$ 
& 0.158 & 0.152 & 0.342 & 0.378 \\ \hline
$\overline{\Sigma^+} (\overline u \overline u \overline s)$ 
& 0.211 & 0.275 & 0.414 & 0.420 \\ \hline
$\overline{\Xi^0}(\overline u \overline s \overline s)$ 
& 0.165 & 0.175 & 0.395 & 0.425 \\ \hline
$\overline{\Xi^-} (\overline d \overline s \overline s)$ 
& 0.152 & 0.126 & 0.274 & 0.381 \\ \hline
$\overline{\Omega} (\overline s \overline s \overline s)$ 
& 0.150 & 0.120 & 0.199 & 0.384 \\ \hline
\end{tabular}
\end{center}
\caption[]{The average $x_F$ for strange hadrons produced at $x_F > 0$ in 
$\pi^- p$ collisions at 500 GeV for all cases considered.
The average $x_F$ for $H$ alone is 0.117.}
\label{avexfdistpi}
\end{table}

In Table~\ref{avexfdistpi}, we give $\langle x_F \rangle$ for all
particles and antiparticles  at $x_F >0$.  
Generally, antistrange hadrons have a larger $\langle x_F \rangle$ than their
corresponding strange hadron.  This occurs because the strange hadron 
distribution at leading twist is larger by a uniform 30\% over all $x_F$.
The multiplication makes the leading twist hadron distribution higher at
low $x_F$ and the antihadron $x_F$ distributions do not fall steeply enough
when fragmentation is included for the average $x_F$ of strange particles to
be larger than those for antistrange particles.
The $\Xi^-$ is an exception because it is leading relative to the 
$\overline{\Xi^-}$.  All the others are either equally leading or nonleading.

In general, $\langle x_F \rangle$ increases between the cases with $H+F+C$ and
$H+C$.  However, for $\Sigma^+$, $\Xi^0$, $\Omega$, $\overline{ \Sigma^-}$,
$\overline{\Xi^-}$, and $\overline \Omega$, $\langle x_F \rangle$ without
fragmentation is less than that with $H+F+C$.  The drop occurs because these
particles are all produced by coalescence only in $n=8$ states so that
eliminating the fragmentation contribution considerably reduces the overall
probability, thus reducing $\langle x_F \rangle$, compare Figs.~\ref{pitot}
and \ref{pitotco}.  Rescaling the relative cross sections, $H+aC$ with $a=40$,
generally
increases $\langle x_F \rangle$ considerably, from 20\% for $K^-$ and $K^0$ to
more than a factor of two for the states first produced when $n=8$, as could
be expected by the existence of a second peak in some of the distributions.
The increase in $\langle x_F \rangle$ between $H+aC$ and $C$ alone is quite
small in some cases, particularly for particles produced in Fock states with
$n=4$ and 6.  The averages for $K^-$ and $\overline{K^0}$, the only hadrons
produced from $n=4$ states, increase by less than 1\%.  Hadrons first produced
with $n=6$, $\overline{K^0}$, $K^+$, $\Lambda$, $\overline \Lambda$,
$\Sigma^-$, $\overline{\Sigma^+}$, $\Xi^-$, and $\overline{\Xi^0}$, show an
increase in $\langle x_F \rangle$ of 2-10\%.  The other hadrons, produced only
when $n=8$ increase somewhat more.  The $\Omega$ and $\overline \Omega$ show
an exceptionally large increase in $\langle x_F \rangle$, a factor of two,
between $H+aC$ and $C$ only.  This large effect is because even when the
intrinsic contribution is enhanced by a factor of 40 the leading-twist cross
section is still dominant.  Also, comparison of the $C$ only results for
$\langle x_F \rangle$ of $\Omega$ and $\overline \Omega$ show that they agree
with $\langle x_F \rangle$ of the $|\overline u d s \overline s s \overline s s
\overline s \rangle$ state in Table~\ref{avexfpi}, as expected.  The same is
true for other hadrons only produced from a single $n=8$ state.  Possible
small differences may arise because while the intrinsic probability is
energy-independent, the cross section is proportional to $\sigma_{hN}^{\rm
in}(\sqrt{s'})$ evaluated at $\sqrt{s}(1-|x_F|)$.

\begin{figure}[htpb]
\setlength{\epsfxsize=0.95\textwidth}
\setlength{\epsfysize=0.5\textheight}
\centerline{\epsffile{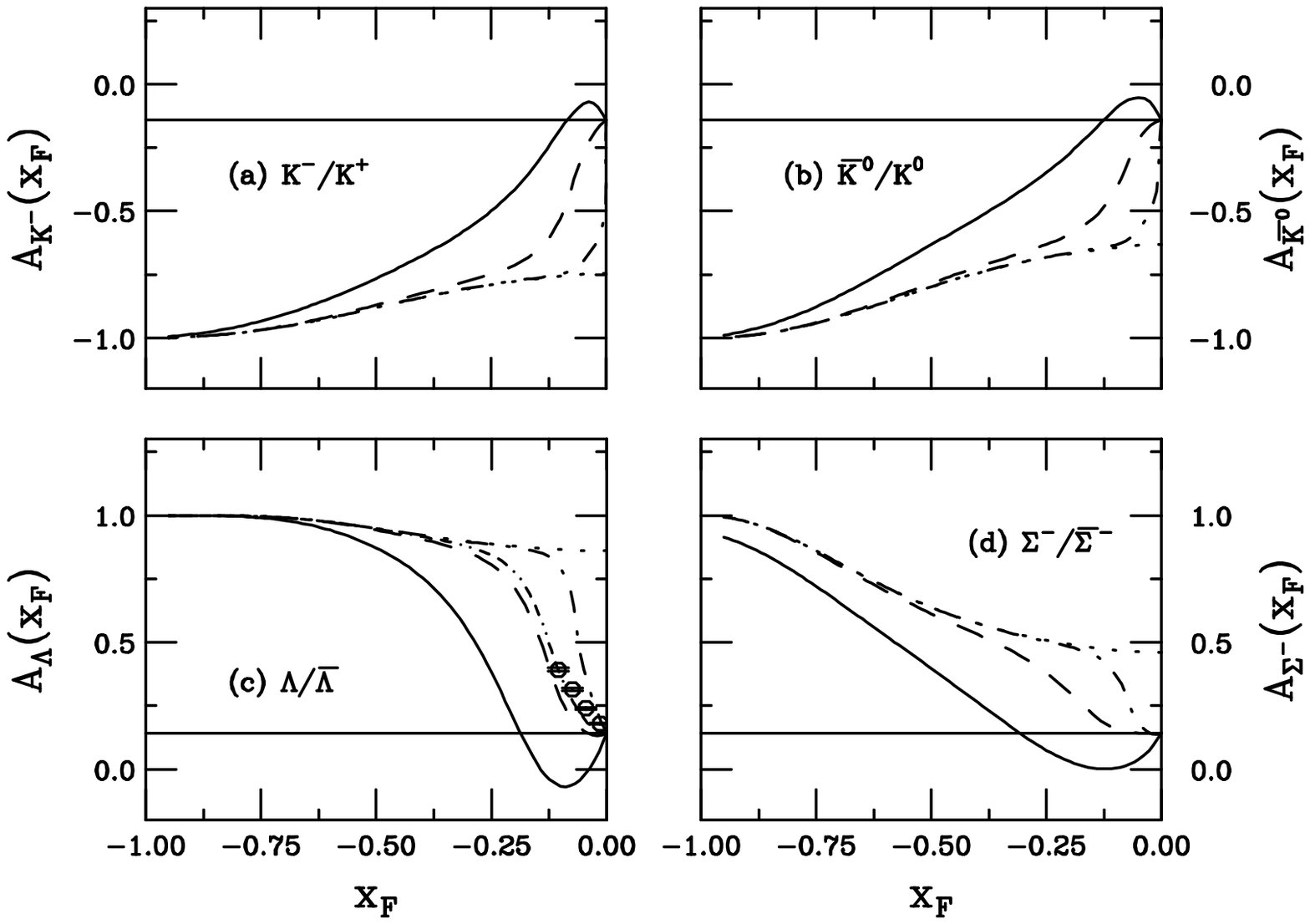}}
\caption[]{Model asymmetries for $\pi^- p$ interactions at 500 GeV with
$H+F+C$ (solid), $H+C$ (dashed), $H+aC$ with $a=40$
(dot-dashed) and $C$ alone (dotted) are shown 
for (a) $A_{K^-}$, (b)
$A_{\overline{K^0}}$, (c) $A_{\Lambda}$ and 
(d) $A_{\Sigma^-}$.  The result with the modified $\overline{\Lambda}$
distributions is shown in the dot-dot-dash-dashed curve in (c).
The horizontal solid curve 
is the asymmetry from leading-twist
production alone. The E791 data \cite{anjosmex1,anjosmex2} 
on $A_\Lambda$ are also shown.  }
\label{apinxf1}
\end{figure}

\begin{figure}[htpb]
\setlength{\epsfxsize=0.95\textwidth}
\setlength{\epsfysize=0.5\textheight}
\centerline{\epsffile{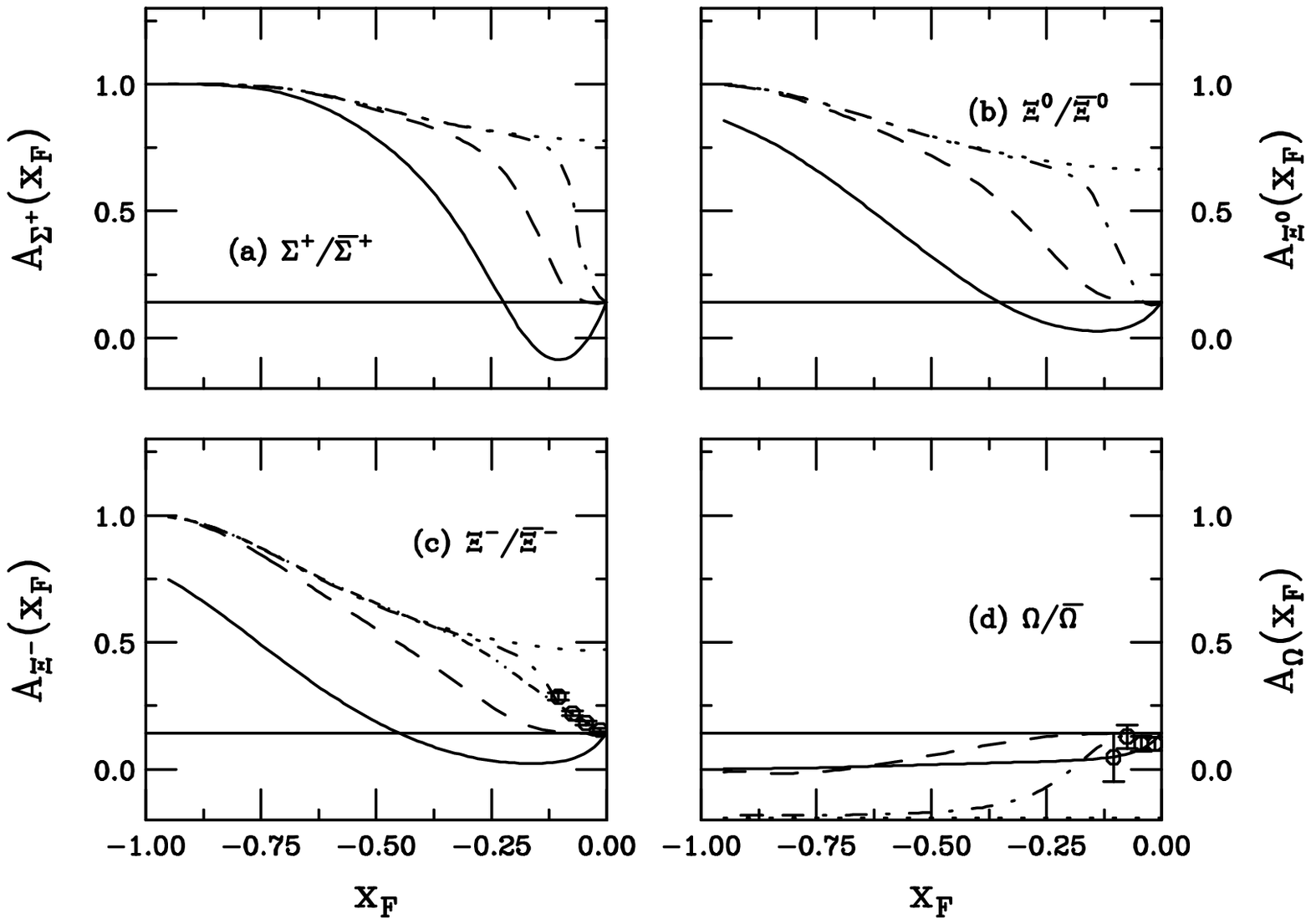}}
\caption[]{ Model asymmetries for $\pi^- p$ interactions at 500 GeV with 
$H+F+C$ (solid), $H+C$ (dashed), $H+aC$ with $a=40$
(dot-dashed) and $C$ alone (dotted) are shown 
for (a) $A_{\Sigma^+}$, (b) 
$A_{\Xi^0}$, (c) $A_{\Xi^-}$ and (d) 
$A_\Omega$.  The result with the modified $\overline{\Xi^-}$
distributions is shown in the dot-dot-dash-dashed curve in (c).
The horizontal solid curve 
is the asymmetry from leading-twist
production alone. The E791 $A_{\Xi^-}$ and $A_\Omega$ 
data \cite{anjosmex1,anjosmex2} 
are also shown.}
\label{apinxf2}
\end{figure}

In Figs.~\ref{apinxf1} and \ref{apinxf2}, we show the corresponding model
calculations at negative $x_F$.  Now all the baryon asymmetries are positive
at large $x_F$ except $A_\Omega$ which is negligible, as might
be expected from the target fragmentation region, typically a proton in a light
target.  The fastest increase in the asymmetry is for $A_{\Lambda}$ and
$A_{\Sigma^+}$ which both have two valence quarks in common
with the proton.  The next largest
strange/antistrange baryon asymmetry is $A_{\Sigma^-}$ because the $\Sigma^-$ 
has a single strange quark
and shares only one valence quark with the proton.  The doubly strange baryon
asymmetries are somewhat weaker with $A_{\Xi^0} > A_{\Xi^-}$ because $\Xi^0$
coalescence production is more probable since the proton has two $u$ valence
quarks and only one $d$ valence quark.  This is the same reason why 
$|A_{K^-}|$ is somewhat larger than $|A_{\overline{K^0}}|$ 
at intermediate $x_F$.  The meson asymmetries are both negative, again by the
definition of $A_S$ as the difference between hadrons with $s$ and $\overline
s$ quarks:  the $K^+$ and $K^0$ cross sections, with $u$ and $d$ quarks in
common with the proton, are larger, changing the sign of $A_S$.

Turning fragmentation off is shown to significantly increase the asymmetries
at low $|x_F|$, even more so than in the pion fragmentation region in 
Figs.~\ref{api1} and \ref{api2}.  The effect is particularly strong for 
$A_\Lambda$ where the ``dip'' due to probability conservation for fragmentation
and coalescence causes the asymmetry to become negative for $|x_F| < 0.15$.
The dip disappears when fragmentation is turned off and $A_\Lambda$ increases
rapidly already at $|x_F| \geq 0$ 
since the $\Lambda$ is produced by coalescence
already in the $|uud s \overline s \rangle$ state.  However, this rapid turn on
still does not increase $A_\Lambda$ as quickly as the data.  The same slower
turn on in $A_{\Xi^-}$ without fragmentation seen in Fig.~\ref{api2}(c) is
also seen here.  

Note also that for $H+C$, at large $|x_F|$ all $A_S \rightarrow 1$ 
except $A_\Omega$ since fragmentation not only builds up the low
to moderate $x_F$ intrinsic contribution
but also tends to mask the effects of coalescence in higher
Fock states.  Thus $A_{\Xi^-} < 1$ at
large $|x_F|$ with $H+F+C$ but $A_{\Xi^-} 
\sim 1$ at high $|x_F|$ for $H+C$ because the $\Xi^-$
is already produced in the seven-particle $|uud s \overline s s \overline s 
\rangle$ state while the $\overline{\Xi^-}$ is only produced in the 
nine-particle $|uud s \overline s s \overline s d \overline d \rangle$ state.
When both strange hadrons are only produced from the same Fock 
state, as is the case for the $\Omega$ and $\overline \Omega$, the asymmetry
is small and nearly independent of $x_F$ since only the relative weight factors
are different.

Similar to the $x_F>0$ results 
shown in Figs.~\ref{api1} and \ref{api2}, 
these calculations reflect the trends of the data but do not increase fast
enough to reproduce it in detail, even for $H+C$.  
Therefore we have also calculated the asymmetries with the same scale factor,
$a=40$, used to fit $A_{\Xi^-}$ in Fig.~\ref{api2}.  While the agreement with 
$A_{\Xi^-}$ in Fig.~\ref{apinxf2} is also quite good, the scaled result
significantly overestimates $A_\Lambda$.  The scaled results converge to those
with $C$ only even more rapidly than at forward $x_F$ since the decrease in the
leading twist cross section is more rapid due to the steeper gluon distribution
in the proton.  The exception is for $A_\Omega$ where the different weights for
$\Omega$ and $\overline \Omega$ result in an overall negative asymmetry.
The different weights also account for the fact that $A_{\Sigma^+} >
A_{\Sigma^-}$ and $A_{\Xi^0} > A_{\Xi^-}$ at $x_F \sim 0$ with $C$ alone.
The $\Sigma^+$ and $\Xi^0$ have greater weights since they share $u$ valence 
quarks with the proton.

\begin{table}
\begin{center}
\begin{tabular}{|c|c|c|c|c|} \hline
Final State & $H+F+C$ & $H+C$ & $H+aC$ & $C$ only \\ \hline
$K^- (\overline u s)$ 
& -0.142 & -0.152 & -0.248 & -0.254 \\ \hline
$\overline{K^0} (\overline d s)$ 
& -0.142 & -0.152 & -0.248 & -0.254 \\ \hline
$\Lambda (uds)$ 
& -0.303 & -0.372 & -0.433 & -0.435 \\ \hline
$\Sigma^- (dds)$
& -0.153 & -0.190 & -0.357 & -0.368 \\ \hline
$\Sigma^+(uus)$ 
& -0.254 & -0.332 & -0.434 & -0.427 \\ \hline
$\Xi^0(uss)$ 
& -0.134 & -0.140 & -0.350 & -0.377 \\ \hline
$\Xi^-(dss)$ 
& -0.128 & -0.116 & -0.329 & -0.374 \\ \hline
$\Omega(sss)$ 
& -0.120 & -0.082 & -0.137 & -0.343 \\ \hline \hline
$K^+ (u \overline s)$  
& -0.222 & -0.258 & -0.292 & -0.293 \\ \hline
$K^0 (d \overline s)$ 
& -0.196 & -0.236 & -0.285 & -0.287 \\ \hline
$\overline \Lambda (\overline u \overline d \overline s)$ 
& -0.134 & -0.128 & -0.311 & -0.337 \\ \hline
$\overline{\Sigma^-} (\overline d \overline d \overline s)$ 
& -0.134 & -0.128 & -0.311 & -0.337 \\ \hline
$\overline{\Sigma^+} (\overline u \overline u \overline s)$ 
& -0.134 & -0.128 & -0.311 & -0.337 \\ \hline
$\overline{\Xi^0}(\overline u \overline s \overline s)$ 
& -0.127 & -0.093 & -0.252 & -0.340 \\ \hline
$\overline{\Xi^-} (\overline d \overline s \overline s)$ 
& -0.127 & -0.093 & -0.252 & -0.340 \\ \hline
$\overline{\Omega} (\overline s \overline s \overline s)$ 
& -0.126 & -0.084 & -0.172 & -0.343 \\ \hline
\end{tabular}
\end{center}
\caption[]{The average $x_F$ for strange hadrons produced at $x_F < 0$ in 
$\pi^- p$ collisions at 500 GeV in all our scenarios.  The average $x_F$ 
for $H$ alone is -0.081.}
\label{avexfdistpinxf}
\end{table}

The average $x_F$ values of the $\pi^- p$ calculations at negative $x_F$
are given in Table~\ref{avexfdistpinxf}.  
The absolute values of these averages are generally
smaller than those in Table~\ref{avexfdistpi}.  This is due in part to the more
steeply falling $x_F$ distribution at $x_F < 0$ in Fig.~\ref{fus} which 
reflects the gluon distribution in the proton.  The nonleading averages
agree according to Eq.~(\ref{proequiv}), as expected.

The data seem to indicate that uncorrelated fragmentation from the intrinsic
state is ruled out, possibly due to the energy cost from a nearly on-shell Fock
state.  Not only is fragmentation ruled out, but the increase in $A_{\Xi^-}$
seems to require a very strong coalescence contribution relative to the
leading-twist result. We now examine this possibility in more detail. 

While we have rescaled the intrinsic contribution, it is not necessarily clear
which component should indeed be changed.  Obviously the intrinsic probability
cannot be increased by such a large factor--it would clearly exceed the
probability sum, $P=1$, in Eq.~(\ref{probsum}).  The only other parameter in
the intrinsic calculation is $\mu^2$, the scale at which the coherence of the
Fock state is broken.  Increasing $\mu^2$ by a factor of 40 would give an
unacceptably large intrinsic cross section, $\sigma_{\rm is} = 9$ and 13 mb for
$\pi^-$ and $p$ interactions respectively, $\sim 40$\% of the inelastic cross
section.  Such a large cross section seems unlikely.
On the leading-twist side, the parameter $B_S$ in the fragmentation function,
Eq.~(\ref{fusfrag}), could be different for different final states.  We assumed
$B_S = 0.1$ for all final states.  However, as discussed previously, the
relative rates of $K$, $\Lambda$, $\Sigma$, $\Xi$ and $\Omega$
production are unknown.  It could well be that $B_S$ should be considerably
smaller for doubly-strange baryons than we have chosen.  Taking a slightly
larger $B_S$ for $\Lambda$ production could account for the overestimate of
$A_\Lambda$ with rescaling.

\begin{figure}[htpb]
\setlength{\epsfxsize=0.95\textwidth}
\setlength{\epsfysize=0.5\textheight}
\centerline{\epsffile{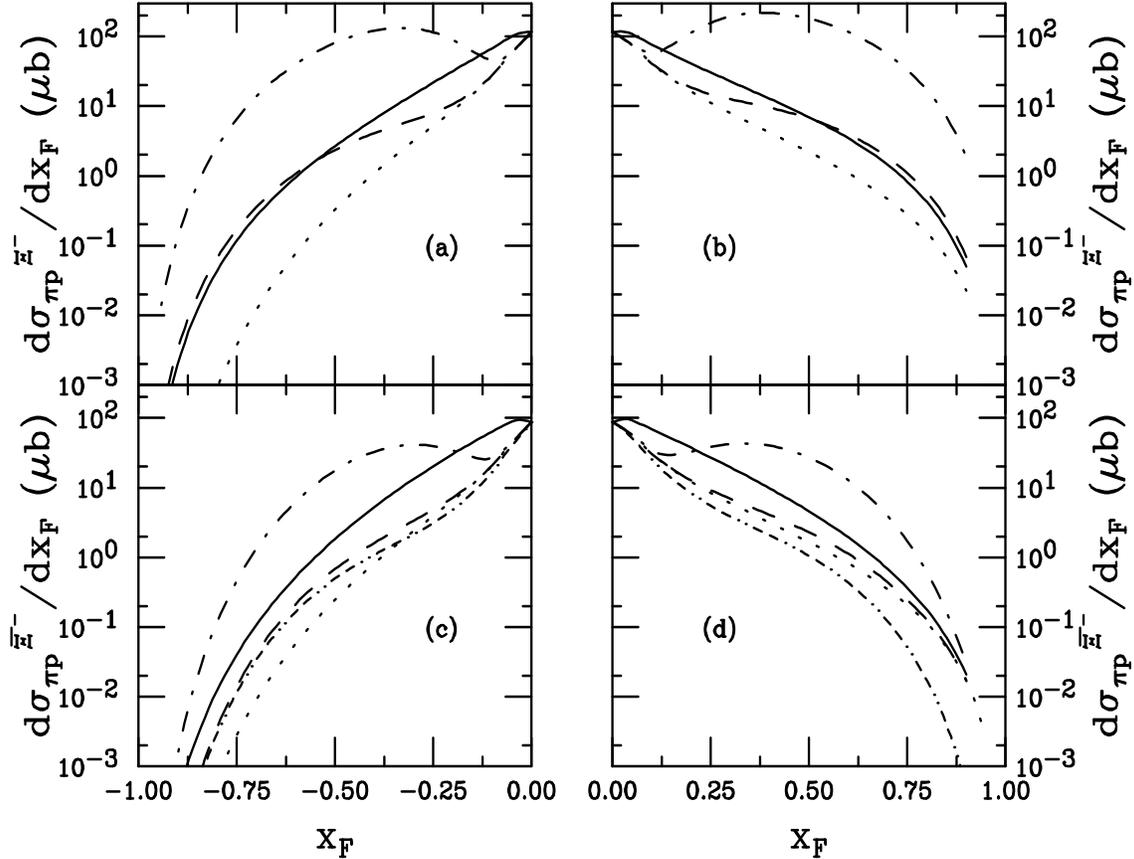}}
\caption[]{The model $\Xi^-$ $x_F$ distributions in the proton (a) and pion (b)
fragmentation regions and the $\overline{\Xi^-}$ $x_F$ distributions in the
proton (c) and pion (d).
The curves show the results for
$H+F+C$ (solid), $H+C$ (dashed), $H+aC$ with $a=40$
(dot-dashed) and $H$ alone (dotted) are shown 
for both $\Xi^-$ and $\overline{\Xi^-}$.  The modified $\overline{\Xi^-}$
distributions, $H' +C$, 
are shown in the dot-dot-dash-dashed curves in (c) and (d).}
\label{ximdists}
\end{figure}

The asymmetries alone do not provide enough information about the individual
cross sections.  A comparison of the inclusive $x_F$ distributions with the
model over a broader range of $x_F$ is essential to check whether there is any
indication of a second peak at intermediate $x_F$.  In Fig.~\ref{ximdists}
we show the individual $x_F$ distributions that are used in the asymmetry
calculations in each case.  There is data on $\Xi^-$ production by pion and
neutron beams on nuclear targets \cite{Biagi,WA89hyp} at lower energies that
are consistent with the relatively small intrinsic component in
Eq.~(\ref{pis}).  In fact, the shapes of the measured $x_F$ distributions 
agree with both the $H+F+C$ and $H+C$ results, showing no evidence for an
enhancement at intermediate $x_F$. 

However, rescaling, as in $H+aC$, is inconsistent with measured $\Xi^-$
distributions in $\pi^- A$ interactions at 345 GeV 
over a larger range of $x_F$ than that covered by the
asymmetries.  Thus we have also checked another 
possibility--that the leading-twist
distribution of strange particles is modified for those hadrons not sharing at
least one valence quark with the ``projectile'' (the pion at $x_F>0$ and the
proton at $x_F <0$).  This modification would retain the agreement
of the model with the $\Xi^-$ $x_F$ distributions while modifying the
asymmetries.  At $x_F > 0$, $A_{\Xi^-}$ would be affected but because both the
$\Omega$ and $\overline \Omega$ distributions would be modified simultaneously,
leaving $A_\Omega$ unchanged.  $A_\Lambda$ would also be unaffected since both
the $\Lambda$ and $\overline \Lambda$ share a valence quark with the $\pi^-$.
The other asymmetries in the forward direction would change.
In the negative $x_F$ region, all the
asymmetries would be affected except $A_\Omega$.  

The assumption that the
leading-twist cross section for hadrons not sharing a valence quark with the
``projectile'' is modified so that $H \rightarrow H' = H(1-x_F)^2$ agrees very
well with $A_{\Xi^-}$ in Figs.~\ref{api2} and \ref{apinxf2} and with
$A_\Lambda$ in Fig.~\ref{apinxf1}, as shown in the dot-dot-dash-dashed curves
for these asymmetries.
With this assumption, $A_{\Xi^-}$ rises more slowly than the $H+aC$
calculations but follows the asymptotic behavior of $H+aC$ and $C$ only at
similar values of $x_F$.  In Fig.~\ref{ximdists}(c), the difference between
the $H+C$ and $H'+C$ curves is rather small for $|x_F| < 0.5$ where the
intrinsic contribution dominates.  At low $|x_F|$, the $H' +C$ distribution
lies below the dotted curve with $H$ alone and it is in this region where the
rapid growth of the asymmetry takes place.  In Fig.~\ref{ximdists}(d), the hard
leading-twist distribution is only slightly below the calculated $H+C$ result
so that the modification $H \rightarrow H'$ puts the $H'+C$ curve lower than
that with $H$ alone over the entire $x_F$ range.

The modification $H \rightarrow H'$ would essentially imply a
modification of the fragmentation function $D_{S/s}$. The difference between
the fragmentation functions of $S$ and $\overline S$ could suggest a breakdown
of factorization.  This may not be surprising given the ``lightness'' of the
strange quark.  The case for perturbative production of strangeness is rather
weak so that it is difficult to rule out such a difference.  
To either verify or disprove the possibility, 
the $x_F$ distribution of the antistrange hadrons
should be measured with sufficient accuracy.

While a modification of the leading-twist distribution for ``nonleading''
hadrons is fairly {\it ad hoc} 
it is the one scenario that agrees with the available
strangeness data \cite{anjosmex1,anjosmex2,Biagi,WA89hyp}.  Therefore any
strong enhancement of the intrinsic contribution is ruled out.  However, the
absence of an intrinsic contribution seems to be ruled out also since, in such
a case, $A_\Lambda = A_{\Xi^-}$ at $x_F<0$, clearly incompatible with the data.
There is no evidence for modification of the leading-twist charm distributions
since the $D^+$ $x_F$ distribution in $\pi^- N$ interactions is consistent with
perturbative QCD.  On the other hand, it is 
interesting to note that the abscence of fragmentation
would actually improve the agreement of the charm asymmetry data with the model
calculations.  

\begin{figure}[htpb]
\setlength{\epsfxsize=0.95\textwidth}
\setlength{\epsfysize=0.5\textheight}
\centerline{\epsffile{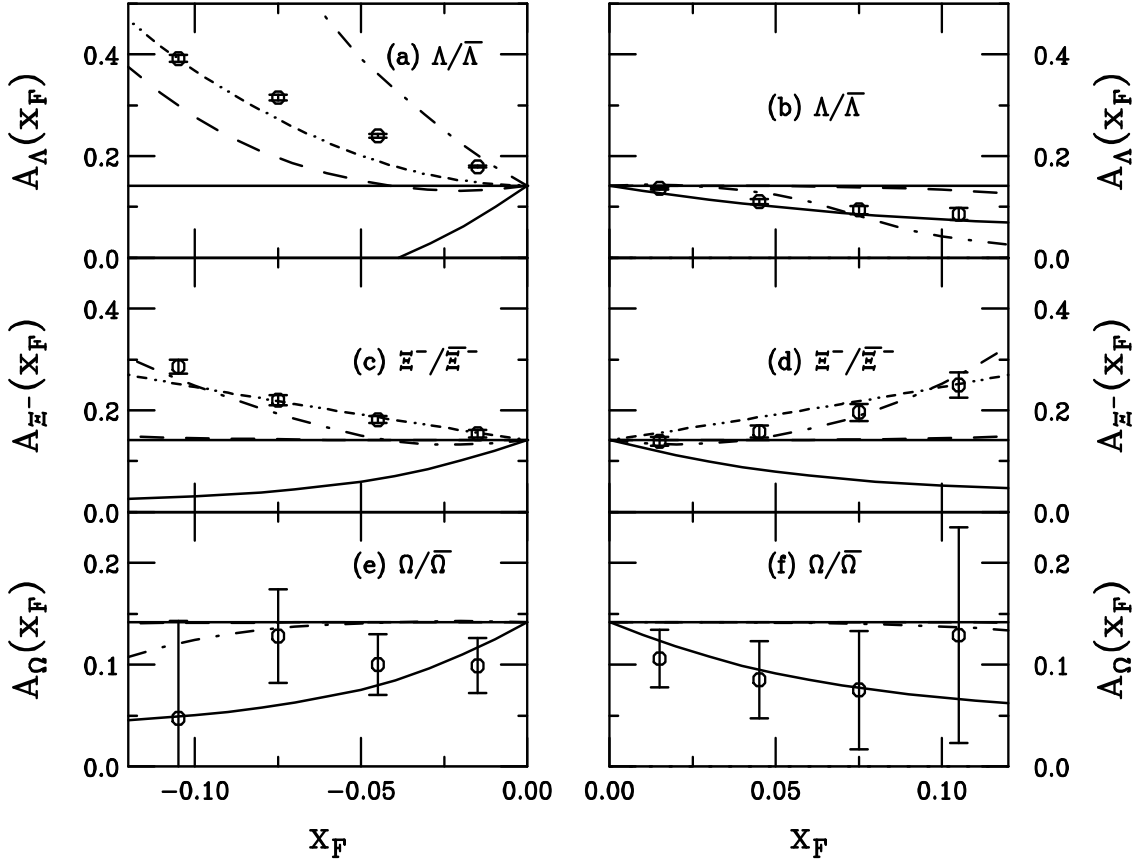}}
\caption[]{The model asymmetries are compared to the data within the range
$|x_F|< 0.1$.  The left-hand-side shows the proton fragmentation
region while the right-hand-side shows the pion fragmentation region.
The curves show the results with
$H+F+C$ (solid), $H+C$ (dashed), $H+aC$ with $a=40$
(dot-dashed) and $H' + C$ (dot-dot-dash-dashed).  The E791 data
\cite{anjosmex1,anjosmex2} are also shown.}
\label{loxf}
\end{figure}

In Fig.~\ref{loxf} we compare our calculations of $A_\Lambda$, $A_{\Xi^-}$
and $A_\Omega$ for the $x_F$ range of the data, $|x_F|\leq 0.12$.  This
enhancement of the low $x_F$ region confirms that the best agreement in
Figs.~\ref{loxf}(a), (c) and (d) is with the modified leading-twist
distribution.  For the other asymmetries in (b), (e) and (f), small
discrepancies could be removed by tuning the results at $x_F = 0$.

Finally, we point out that we have assumed proton targets in all cases.  
If we assume a nuclear target and take into account
both neutrons and protons in the target fragmentation region, the differences
between, for example, $A_{K^-}$ and $A_{\overline{K^0}}$ would disappear 
if the target had an equal number of neutrons and protons.  In addition, the
intrinsic model predicts that the $A$ dependence should be weaker than linear,
$A^{0.71}$ for protons and $A^{0.77}$ for pions \cite{VBH1}.  Thus the 
asymmetries would decrease at intermediate values of $x_F$ for nuclear targets.
We will study strange particle production, particularly of $\Xi^-$ and
$\Omega$, as a function of $x_F$ 
on nuclear targets \cite{Biagi,WA89hyp} by a variety of projectiles 
in a future work \cite{GutVogtip}.

\section{Summary and Conclusions}

We have extended the intrinsic charm model of Refs.\
\cite{VB,VBlam,VBH2} to strange hadrons.  We have inferred the probabilities
for the Fock states with 1-3 intrinsic $Q \overline Q$ pairs.
We calculated the strange hadron
distributions predicted in the intrinsic model for $\pi^- p$
interactions.  We find that the model predicts asymmetries at
lower values of $x_F$ than for the more massive charm quarks.
We correctly produce the general
trends of the $\pi^- p$ data but not the strong increase of the 
asymmetry at low $|x_F|$,
even when intrinsic-model fragmentation is switched
off.  The  data are suggestive that fragmentation is not effective in the
intrinsic model.  The increase in the asymmetries $A_{\Xi^-}$ and $A_\Lambda$
with $x_F$ cannot be reproduced in the model unless either the intrinsic cross
section is increased greatly or the shape of the leading-twist nonleading
distribution is modified.  Increasing the intrinsic cross section to obtain
agreement with the asymmetries modifies the individual $x_F$ distributions too
strongly, destroying agreement with inclusive $\Xi^-$ spectra
\cite{Biagi,WA89hyp}.  Modifying the leading-twist distribution is consistent
with all data so far but the inclusive $x_F$ distributions of
$\overline{\Xi^-}$ are unavailable.  We have also shown that the modified
leading-twist distribution alone cannot describe the asymmetries since then
$A_\Lambda$ and $A_{\Xi^-}$ should then be identical in the proton
fragmentation region.  Precision data are clearly needed, particularly on the
antistrange baryon $x_F$ distributions, to test these hypotheses.

Acknowledgments: We thank J.C. Anjos, S.J. Brodsky, J. Engelfried, G. Herrera, 
P. Hoyer, E. Ramberg and J. Rathsman for discussions.  R.V. would like to thank
the Geschellschaft f\"{u}r Schwerionenforschung and the Niels Bohr Institute
for hospitality during the completion of this work.

\newpage
\setcounter{equation}{0}
\renewcommand{\theequation}{A.\arabic{equation}}
\begin{center}
{\bf Appendix A}
\end{center}
\vspace{0.2in}

\begin{figure}[htpb]
\setlength{\epsfxsize=0.95\textwidth}
\setlength{\epsfysize=0.5\textheight}
\centerline{\epsffile{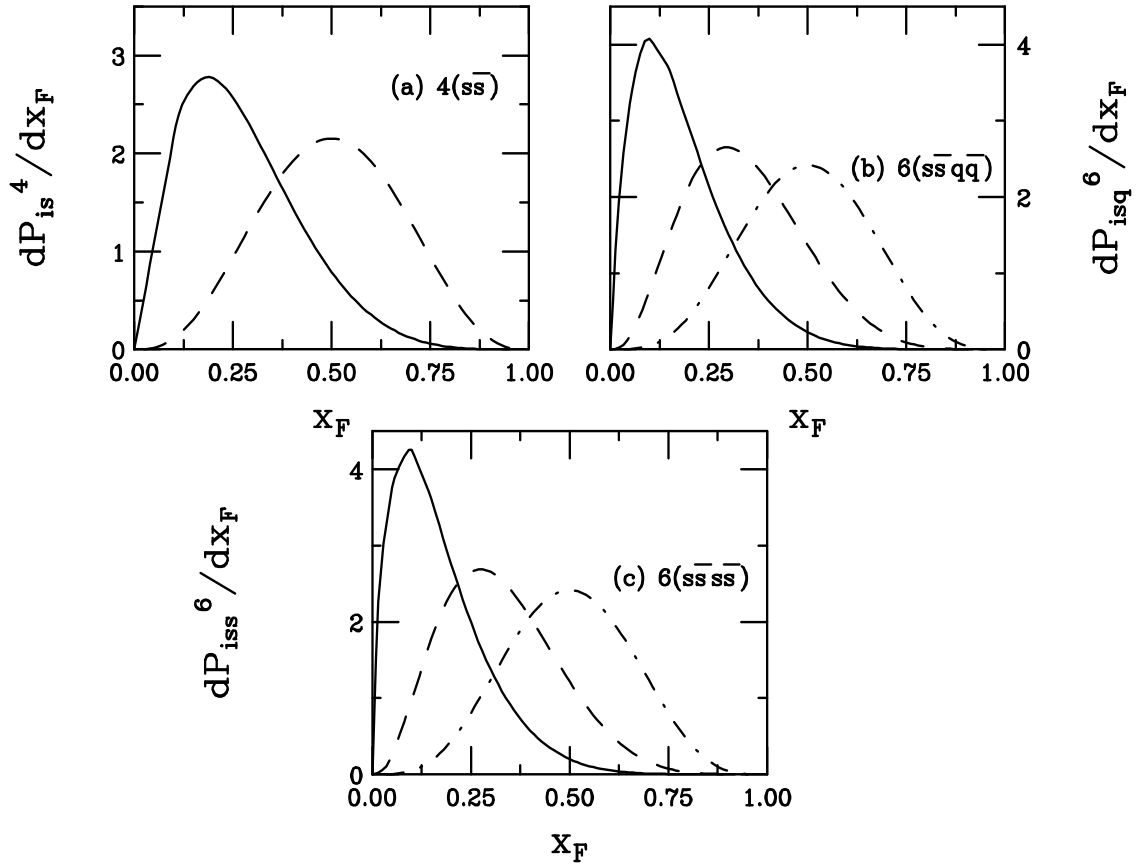}}
\caption[]{ Strange hadron production in the intrinsic model from a
$\pi^-$ projectile in a minimal 4-particle $s \overline s$ Fock state, (a),
a 6-particle Fock 
state with light quark pair, $q=u$ or $d$, and one $s \overline s$ pair, 
(b), and with 
two $s \overline s$ pairs (c).  Both the uncorrelated fragmentation and
coalescence distributions are shown.  The solid
curve in each case is the strange quark distribution, equivalent to the
hadron distribution from uncorrelated fragmentation.  The other curves are
the probability distributions for hadron production by coalescence. The dashed
curves are the $K$ meson distributions.  
The dot-dashed curve in (b) is the baryon 
or antibaryon distribution with a single $s/\overline s$ quark while the 
dot-dashed curve in (c) is the doubly-strange baryon/antibaryon distribution.
}
\label{probpi46}
\end{figure}

\begin{figure}[htpb]
\setlength{\epsfxsize=0.95\textwidth}
\setlength{\epsfysize=0.5\textheight}
\centerline{\epsffile{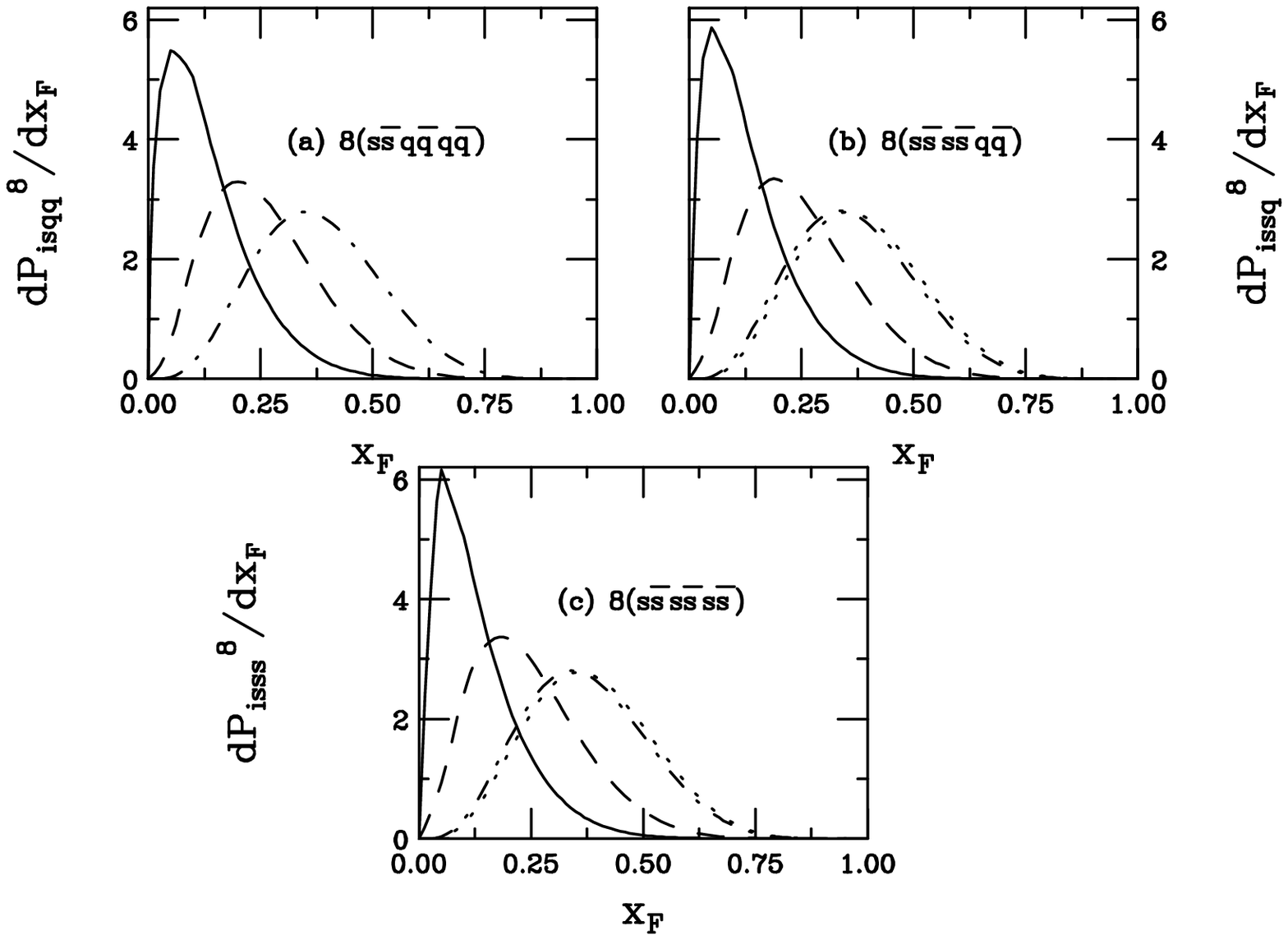}}
\caption[]{ Strange hadron production in the intrinsic model from a
$\pi^-$ projectile in 8-particle Fock states with one, (a), two, (b), and three
(c) $s \overline s$ pairs.  The light quark pairs, 
denoted $q$, refer to both $u$ 
and $d$ quarks.  Both the uncorrelated fragmentation and
coalescence distributions are shown.  The solid
curve in each case is the strange quark distribution, also the
hadron distribution from uncorrelated fragmentation.  The other curves are
the probability distributions for hadron production by coalescence. The dashed
curves are the $K$ meson distributions.  The dot-dashed curves in (a) and (b)
are baryons or antibaryons with a single $s/\overline s$ quark while the
dot-dashed curve in (c) is the doubly-strange baryon/antibaryon distribution.  
The dotted curves in (b) is
the doubly-strange baryon/antibaryon distributions while the dotted curve in 
(c) is the triply-strange $\Omega/\overline{\Omega}$ distribution.
}
\label{probpi8}
\end{figure}

\begin{figure}[htpb]
\setlength{\epsfxsize=0.95\textwidth}
\setlength{\epsfysize=0.5\textheight}
\centerline{\epsffile{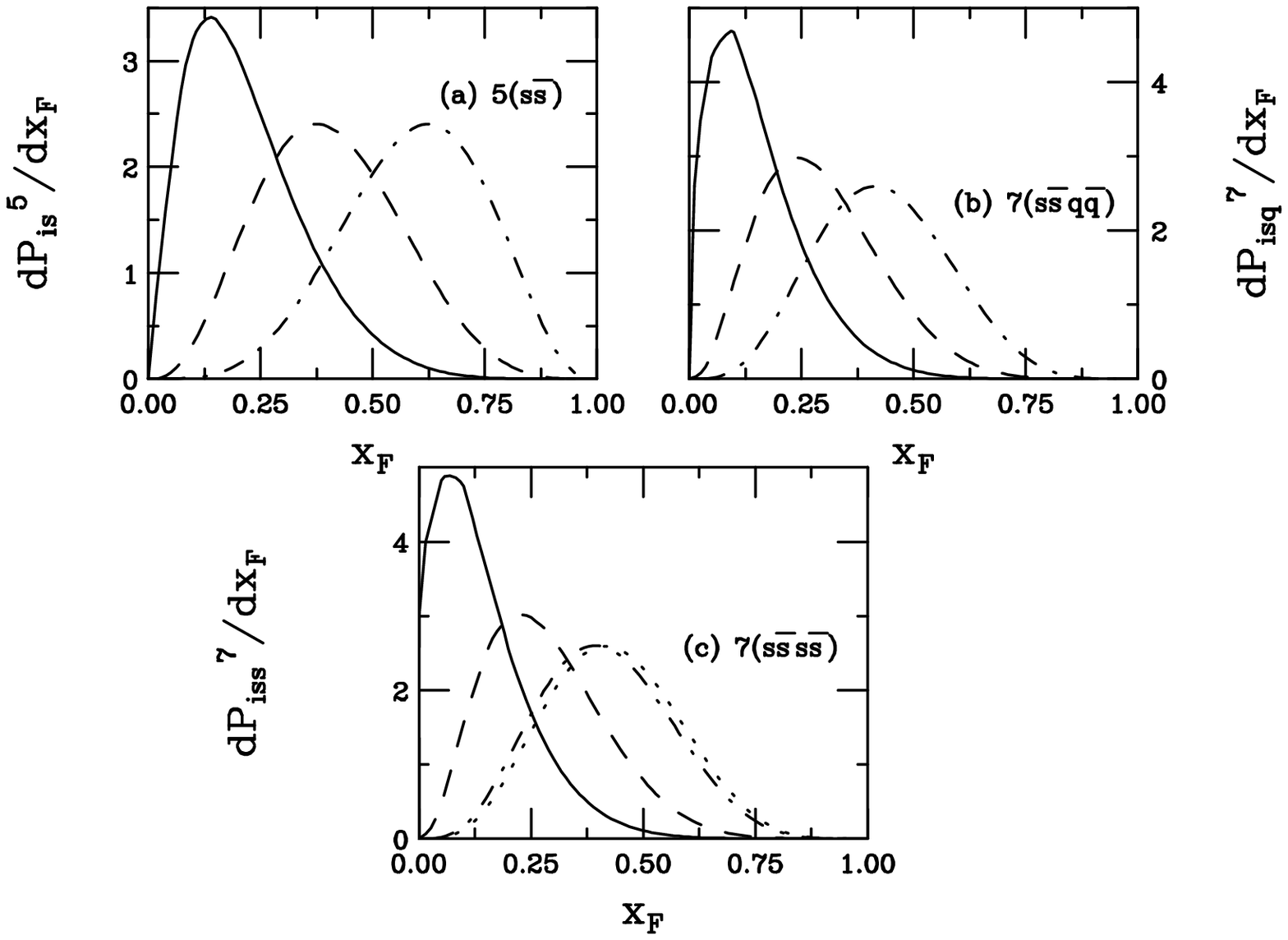}}
\caption[]{ Strange hadron production in the intrinsic model from a
proton projectile in a minimal 5-particle $s \overline s$ Fock state, (a),
a 7-particle Fock 
state with one light quark pair, $q=u$ or $d$, and one $s \overline s$ pairs, 
(b), and with 
two $s \overline s$ pairs (c).  Both the uncorrelated fragmentation and
coalescence distributions are shown.  The solid
curve in each case is the strange quark distribution, also the
hadron distribution from uncorrelated fragmentation.  The other curves are
the probability distributions for hadron production by coalescence. The dashed
curves are the $K$ meson
distributions. The dot-dashed curves are baryons with a
single strange quark, and the dotted curve in (c) is the
doubly strange baryon distribution.
}
\label{probp57}
\end{figure}

\begin{figure}[htpb]
\setlength{\epsfxsize=0.95\textwidth}
\setlength{\epsfysize=0.5\textheight}
\centerline{\epsffile{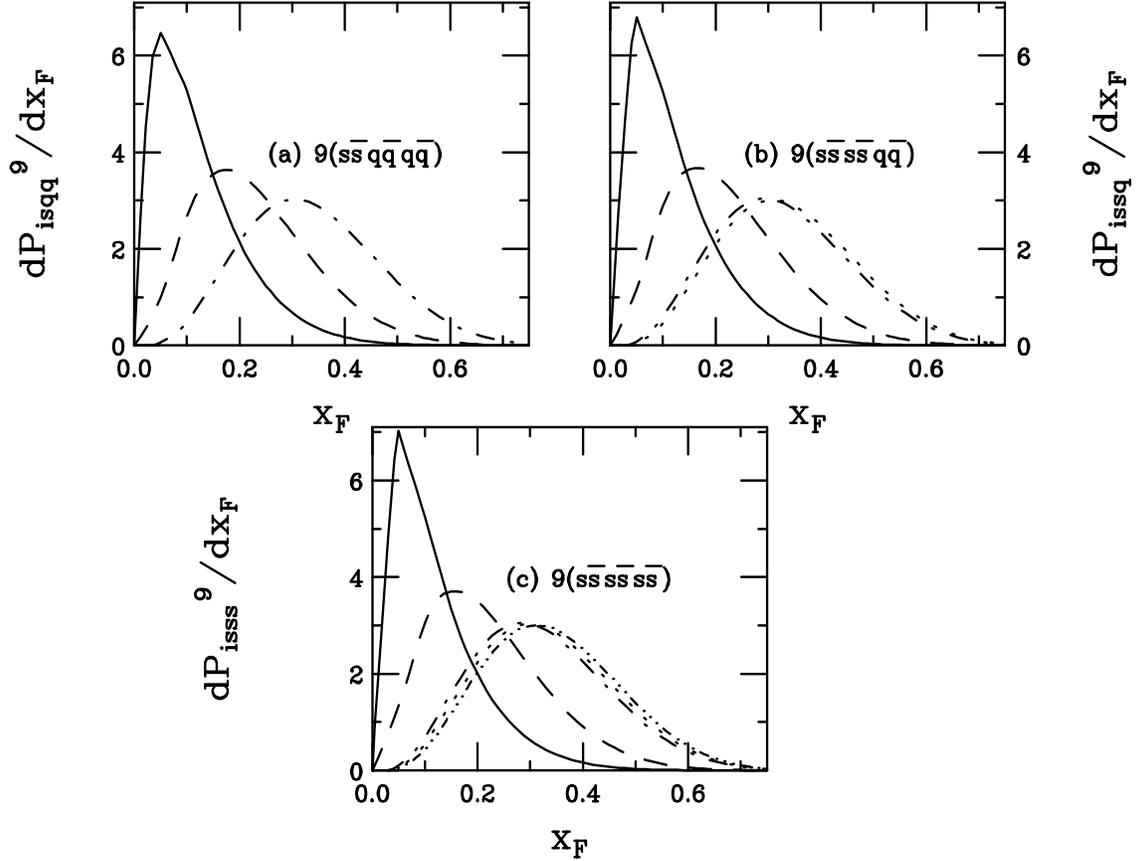}}
\caption[]{ Strange hadron production in the intrinsic model from a
proton projectile in 9-particle Fock states with one, (a), two, (b), and three
(c) $s \overline s$ pairs.  The light quark pairs, denoted $q$, refer 
to both $u$ 
and $d$ quarks.  Both the uncorrelated fragmentation and
coalescence distributions are shown.  The solid
curve in each case is the strange quark distribution, also the
hadron distribution from uncorrelated fragmentation.  The other curves are
the probability distributions for hadron production by coalescence. The dashed
curves are the $K$ meson distributions.  The dot-dashed curves are baryons 
or antibaryons (antibaryons in (a) only) with a single $s/\overline s$ quark, 
the dotted curves are doubly-strange baryons or antibaryons (antibaryons in (b)
only), and the dot-dot-dot-dashed curve in (c)
is the triply-strange $\Omega/\overline{\Omega}$.
}
\label{probp9}
\end{figure}

In this appendix, we show the normalized probability distributions
$(1/P^n_{\rm iQ})(dP^n_{\rm iQ}/dx_F)$, for both uncorrelated fragmentation
and coalescence are given for the
pion and proton Fock states in Figs.~\ref{probpi46}-\ref{probp9}. 
These probability distributions, when properly weighted, 
will comprise the intrinsic contribution to strange hadron production.  
The probability distributions for pions from Fock states with $n=4$, 6 and 8
are given in Figs.~\ref{probpi46} and \ref{probpi8}.  Figures \ref{probp57} and
\ref{probp9} are the corresponding intrinsic probability distributions
from Fock states with $n=5$, 7 and 9.

It is clear from a comparison of the strange quark distributions in
Figs.~\ref{probp57}(a), \ref{probp57}(b), and \ref{probp9}(a) from five-, 
seven-, and nine-parton Fock states
that the strange quark takes less of the projectile momentum as the
number of partons in the configuration increases.  These distributions 
correspond to production of all strange hadrons by uncorrelated fragmentation.
As $n$ increases, the strange quark distribution is suggestive of those in
parameterizations of the parton distribution functions obtained from fits
to data except for the behavior as $x_F \rightarrow 0$.  If still 
more partons were included in the Fock state, the peak of the $x_F$ 
distribution would occur at smaller $x_F$.
The intrinsic model does not distinguish between 
``valence'' and ``sea'' quarks in the state and treats all partons similarly
except for the $\widehat{m}_Q^2$ weighting of each parton momentum fraction.

The coalescence curves are representative only since for 
example, in the dashed curve in Fig.~\ref{probp57}(a), 
the only $K$ mesons produced by coalescence from the
$|uud s \overline s \rangle$ state are $K^+$ and $K^0$.  The probability 
distributions are the same for both mesons before any weight factors are taken
into account, as will 
be described later.  These weights only change the relative 
normalization from a given state, not the shape of the distribution.  In
some of the higher Fock states, all $K$ mesons can be produced by coalescence
and then the meson probability distribution is the same for all kaons even
though the weights are different for each meson.  The
same is true for the $\Lambda$ and $\Sigma^+$ distributions in
the dot-dashed curve in Fig.~\ref{probp57}(a), 
and for all baryons with a single
strange quark in the higher Fock states.  When it is possible to produce 
baryons with more than one strange quark, the average $x_F$ of the 
multiply-strange hadron is the largest of all the hadrons produced by 
coalescence because of the extra momentum imparted by the more massive strange
quarks.  

When eight- or nine-particle states are considered, 
both strange baryons and strange antibaryons can be produced by coalescence.  
In any given eight- or nine-particle Fock state
then, the strange baryon and any antiparticle counterpart have the
same probability distributions although they may have different weight 
factors.  It is
only the fact that, in most cases, the strange baryon can be produced in Fock
states with fewer particles that gives it the ``leading'' edge over the strange
antibaryon.  This is especially true for the $\Omega$ and the 
$\overline{\Omega}$ which have equal probabilities and total intrinsic 
distributions from a pion but in a $\Sigma^-$, the
$\Omega$ can already be produced by coalescence from a 7-particle Fock state
while the $\overline{\Omega}$ is only produced by coalescence in 
the nine-particle
$|dds s \overline s s \overline s s \overline s \rangle $ state. 

The average momentum fractions, $\langle x_F \rangle$, of all these generic 
Fock states are given in Tables~\ref{avexfpi} and \ref{avexfpro}.  The average
momentum fractions carried by the strange quarks decreases $\sim 50$\% for
all projectile hadrons between the minimal and the nine-particle Fock states.  
On average the strange quarks carry more momentum
in the pion because of its lower valence quark content.  Also, in the 
four-parton Fock state, 
the $K^-$ and $K^0$ can take half the pion momentum when
produced by coalescence while the singly and doubly-strange baryons take
half the pion momentum in the $|\overline ud q \overline q s \overline s
\rangle$ and $|\overline u d s \overline s s \overline s \rangle$ six-parton 
Fock states respectively.  In general, the strange baryons take more of 
the momentum from the lower $p$ and $\Sigma^-$ Fock states than from the pion 
while the strange mesons generally take less momentum
from the proton than the pion.
The situation is reversed between the eight-parton states of the pion and the 
nine-parton states of the proton and $\Sigma^-$ because the additional parton
in the projectile baryons dilutes the available momentum for coalescence 
sufficiently to reduce the $\langle x_F \rangle$ with a baryon projectile 
relative to a pion projectile.

\begin{table}
\begin{center}
\begin{tabular}{|ccc|ccc|} \hline
State & Particle & $\langle x_F \rangle$ & State & Particle & 
$\langle x_F \rangle$ \\ \hline
$|\overline u d s \overline s \rangle$ & $s$ & 0.272 & $|\overline u d s 
\overline s q \overline q q \overline q \rangle$ & $s$ & 0.138 \\ \hline
 '' & $ \overline q s = q \overline s$ & 0.500 & '' & $ \overline q s = q 
\overline s$ & 0.258 \\ \hline
$|\overline u d s \overline s q \overline q \rangle$ & $s$ & 0.182 
& '' & $qqs = \overline q \overline q \overline s$ & 0.379 \\ \hline
 '' & $\overline q s = q \overline s$ & 0.342 & $|\overline u d s \overline s 
s \overline s q \overline q \rangle$ & $s$ & 0.133 \\ \hline
 '' & $qqs = \overline q \overline q \overline s$ & 0.500 
& '' & $ \overline q s = q \overline s$ & 0.250 \\ \hline
$|\overline u d s \overline s s \overline s \rangle$ & $s$ & 0.173 
& '' & $qqs = \overline q \overline q \overline s$ & 0.368 \\ \hline
 '' & $\overline q s = q \overline s$ & 0.326 & '' & $qss = \overline q 
\overline s \overline s$ & 0.382 \\ \hline
 '' & $qss = \overline q \overline s \overline s$ & 0.500 & 
$|\overline u d s \overline s s \overline s s \overline s \rangle$ & $s$ 
& 0.130 \\ \hline
 & & & '' & $\overline q s = q \overline s$ & 0.243 \\ \hline
 & & & '' & $qss = \overline q \overline s \overline s$ & 0.371 \\ \hline
 & & & '' & $sss = \overline s \overline s \overline s$ & 0.384 \\ \hline
\end{tabular}
\end{center}
\caption[]{The average value of $x_F$ for strange hadrons produced by 
fragmentation and coalescence from pion projectiles in 4, 6, and 8 parton
configurations with $q = u$ or $d$.}
\label{avexfpi}
\end{table}

\begin{table}
\begin{center}
\begin{tabular}{|ccc|ccc|} \hline
State & Particle & $\langle x_F \rangle$ & State & Particle & 
$\langle x_F \rangle$ \\ \hline
$|uud s \overline s \rangle$ & $s$ & 0.220 & $|uud s \overline s q \overline q 
q \overline q \rangle$ & $s$ & 0.123 \\ \hline
 '' & $ \overline q s = q \overline s$ & 0.407 & '' & $ \overline q s = q 
\overline s$ & 0.230 \\ \hline
 '' & $qqs$ & 0.593 & '' & $qqs = \overline q \overline q \overline s$ & 0.338
\\ \hline
$|uud s \overline s q \overline q \rangle$ & $s$ & 0.157 & $|uud s \overline 
s s \overline s q \overline q \rangle$ & $s$ & 0.120 \\ \hline
 '' & $\overline q s = q \overline s$ & 0.295 & '' & $ \overline q s = q 
\overline s$ & 0.223 \\ \hline 
'' & $qqs$ & 0.432 & '' & $qqs$ & 0.330 \\ \hline
$|uud s \overline s s \overline s \rangle$ & $s$ & 0.150 & '' & $qss =
\overline q \overline s \overline s$ & 0.342 \\ \hline
 '' & $\overline q s = q \overline s$ & 0.283 & $|uud s \overline s s 
\overline s s \overline s \rangle$ & $s$ & 0.118 \\ \hline
 '' & $qqs$ & 0.416 & '' & $\overline q s = q \overline s$ & 0.218 \\ \hline
 '' & $qss$ & 0.434 & '' & $qqs$ & 0.323 \\ \hline
 & & & '' & $qss$ & 0.333 \\ \hline
 & & & '' & $sss = \overline s \overline s \overline s$ & 0.344 \\ \hline
\end{tabular}
\end{center}
\caption[]{The average value of $x_F$ for strange hadrons produced by 
fragmentation and coalescence from proton projectiles in 5, 7, and 9 parton
configurations with $q = u$ or $d$.}
\label{avexfpro}
\end{table}

\newpage
\setcounter{equation}{0}
\renewcommand{\theequation}{B.\arabic{equation}}
\begin{center}
{\bf Appendix B}
\end{center}
\vspace{0.2in}

Here we give the total probability distributions for strange and antistrange
hadron production from the intrinsic model from $\pi^-$ and $p$
projectiles.  Production from all states with up to three additional $Q
\overline Q$ pairs in the Fock configuration is included.  The
probability distributions for each possible final state combination from
uncorrelated fragmentation and coalescence are given in
Figs.~\ref{probpi46}-\ref{probp9}.  We sum all 
the probabilities over the all the states for each projectile to find the total
strange hadron $x_F$ distribution from the intrinsic model.  Thus, from
Eq.~(\ref{intsumap}), we have
\be \frac{dP_S}{dx_F} = \sum_n \sum_{r_u} \sum_{r_d} \sum_{r_s} \beta
\left( \frac{1}{10} \frac{dP_{{\rm i} (r_s {\rm s}) (r_u {\rm u}) (r_d {\rm d})
}^{nF}}{dx_F} + \xi \frac{dP_{{\rm i} (r_s {\rm s}) (r_u {\rm u}) (r_d {\rm d})
}^{nC}}{dx_F} \right)
\label{intsumap}
\ee
The weight of each state produced by coalescence is
$\xi$ where $\xi = 0$ when $S$ cannot be produced by coalescence from state
$|n_v  r_s(s \overline s) r_u(u \overline u)r_d(d \overline d) \rangle$.  
The parameter $\beta$ is 1 when $\xi = 0$ 
and 0.5 when production by both fragmentation and coalescence is 
possible to conserve probability in each state.  
When we assume coalescence production only, $P^{nF} 
\equiv 0$ and $\beta \equiv 1$.
The number of up, down and strange $Q \overline Q$ pairs 
is indicated by $r_u$, $r_d$ and $r_s$ respectively.  The total, $r_u + r_d
+ r_s = r$, is defined as $r = (n - n_v)/2$ because each $Q$ in an $n$-parton
state is accompanied by a $\overline Q$. For baryon projectiles, $n =5$, 7, 
and 9 while for mesons $n=4$, 6, and 8.  Depending on the value of $n$, $r_i$
can be 0, 1, 2 or 3, {\it e.g.}\ in a $|uud s \overline s d \overline d d 
\overline d \rangle$ state, $r_u = 0$, $r_d = 2$ and $r_s = 1$ with $r = 3$.  
We note that the predictions for $\Lambda$
and $\Sigma^0$ are identical because their quark content is the
same. The normalized probabilities for the Fock states with two and three
additional $Q \overline Q$ pairs are given in Eqs.~(\ref{pisq})-(\ref{pisss}).
Recall that $P^5_{\rm is} = 0.02$.

The strange and antistrange hadron probability
distributions from a $\pi^-$ projectile are:
\be
\frac{dP_{K^+}}{dx_F} & = & \frac{1}{10} \frac{dP_{\rm
is}^{4F}}{dx_F} + 
 \frac{1}{2} \left( \frac{1}{10} \frac{dP_{\rm
isu}^{6F}}{dx_F} + \frac{1}{4} \frac{dP_{\rm isu}^{6C}}{dx_F} \right) 
+ \frac{1}{10} \frac{dP_{\rm
isd}^{6F}}{dx_F} + \frac{1}{10} \frac{dP_{\rm
iss}^{6F}}{dx_F} \nonumber
\\ &   & + \, \frac{1}{2} \left( \frac{1}{10} \frac{dP_{\rm
isuu}^{8F}}{dx_F} + \frac{2}{7} \frac{dP_{\rm isuu}^{8C}}{dx_F} \right) +
 \frac{1}{2} \left( \frac{1}{10} \frac{dP_{\rm
isud}^{8F}}{dx_F} 
+ \frac{1}{7} \frac{dP_{\rm isud}^{8C}}{dx_F} \right)\nonumber
\\ &   & + \, \frac{1}{10} \frac{dP_{\rm
isdd}^{8F}}{dx_F} + 
 \frac{1}{2} \left( \frac{1}{10} \frac{dP_{\rm
issu}^{8F}}{dx_F} + \frac{2}{12} \frac{dP_{\rm issu}^{8C}}{dx_F} \right) 
+ \frac{1}{10} \frac{dP_{\rm
issd}^{8F}}{dx_F} + \frac{1}{10} \frac{dP_{\rm
isss}^{8F}}{dx_F} \, \, ,
\label{piprobkp} 
\\
\frac{dP_{K^0}}{dx_F} & = & \frac{1}{2} \left( \frac{1}{10} \frac{dP_{\rm
is}^{4F}}{dx_F} + \frac{1}{2} \frac{dP_{\rm is}^{4C}}{dx_F} \right) + 
 \frac{1}{2} \left( \frac{1}{10} \frac{dP_{\rm
isu}^{6F}}{dx_F} + \frac{1}{4} \frac{dP_{\rm isu}^{6C}}{dx_F} \right) \nonumber
\\ &   & + \, \frac{1}{2} \left( \frac{1}{10} \frac{dP_{\rm
isd}^{6F}}{dx_F} + \frac{2}{4} \frac{dP_{\rm isd}^{6C}}{dx_F} \right) +
 \frac{1}{2} \left( \frac{1}{10} \frac{dP_{\rm
iss}^{6F}}{dx_F} + \frac{2}{7} \frac{dP_{\rm iss}^{6C}}{dx_F} \right) \nonumber
\\ &   & + \, \frac{1}{2} \left( \frac{1}{10} \frac{dP_{\rm
isuu}^{8F}}{dx_F} + \frac{1}{7} \frac{dP_{\rm isuu}^{8C}}{dx_F} \right) +
 \frac{1}{2} \left( \frac{1}{10} \frac{dP_{\rm
isud}^{8F}}{dx_F} 
+ \frac{2}{7} \frac{dP_{\rm isud}^{8C}}{dx_F} \right)\nonumber
\\ &   & + \, \frac{1}{2} \left( \frac{1}{10} \frac{dP_{\rm
isdd}^{8F}}{dx_F} + \frac{3}{7} \frac{dP_{\rm isdd}^{8C}}{dx_F} \right) +
 \frac{1}{2} \left( \frac{1}{10} \frac{dP_{\rm
issu}^{8F}}{dx_F} + \frac{2}{12} \frac{dP_{\rm issu}^{8C}}{dx_F} \right) 
\nonumber \\ &   & + \, \frac{1}{2} \left( \frac{1}{10} \frac{dP_{\rm
issd}^{8F}}{dx_F} + \frac{4}{12} \frac{dP_{\rm issd}^{8C}}{dx_F} \right) +
 \frac{1}{2} \left( \frac{1}{10} \frac{dP_{\rm
isss}^{8F}}{dx_F} + \frac{3}{16} \frac{dP_{\rm isss}^{8C}}{dx_F} \right)\, \, ,
\label{piprobk0} 
\\
\frac{dP_{K^-}}{dx_F} & = & \frac{1}{2} \left( \frac{1}{10} \frac{dP_{\rm
is}^{4F}}{dx_F} + \frac{1}{2} \frac{dP_{\rm is}^{4C}}{dx_F} \right) + 
 \frac{1}{2} \left( \frac{1}{10} \frac{dP_{\rm
isu}^{6F}}{dx_F} + \frac{2}{4} \frac{dP_{\rm isu}^{6C}}{dx_F} \right) \nonumber
\\ &   & + \, \frac{1}{2} \left( \frac{1}{10} \frac{dP_{\rm
isd}^{6F}}{dx_F} + \frac{1}{4} \frac{dP_{\rm isd}^{6C}}{dx_F} \right) +
 \frac{1}{2} \left( \frac{1}{10} \frac{dP_{\rm
iss}^{6F}}{dx_F} + \frac{2}{7} \frac{dP_{\rm iss}^{6C}}{dx_F} \right) \nonumber
\\ &   & + \, \frac{1}{2} \left( \frac{1}{10} \frac{dP_{\rm
isuu}^{8F}}{dx_F} + \frac{3}{7} \frac{dP_{\rm isuu}^{8C}}{dx_F} \right) +
 \frac{1}{2} \left( \frac{1}{10} \frac{dP_{\rm
isud}^{8F}}{dx_F} 
+ \frac{2}{7} \frac{dP_{\rm isud}^{8C}}{dx_F} \right)\nonumber
\\ &   & + \, \frac{1}{2} \left( \frac{1}{10} \frac{dP_{\rm
isdd}^{8F}}{dx_F} + \frac{1}{7} \frac{dP_{\rm isdd}^{8C}}{dx_F} \right) +
 \frac{1}{2} \left( \frac{1}{10} \frac{dP_{\rm
issu}^{8F}}{dx_F} + \frac{4}{12} \frac{dP_{\rm issu}^{8C}}{dx_F} \right) 
\nonumber \\ &   & + \, \frac{1}{2} \left( \frac{1}{10} \frac{dP_{\rm
issd}^{8F}}{dx_F} + \frac{2}{12} \frac{dP_{\rm issd}^{8C}}{dx_F} \right) +
 \frac{1}{2} \left( \frac{1}{10} \frac{dP_{\rm
isss}^{8F}}{dx_F} + \frac{3}{16} \frac{dP_{\rm isss}^{8C}}{dx_F} \right)\,\, , 
\label{piprobkm} 
\\
\frac{dP_{\overline K^0}}{dx_F} & = & \frac{1}{10} \frac{dP_{\rm
is}^{4F}}{dx_F} + \frac{1}{10} \frac{dP_{\rm
isu}^{6F}}{dx_F} + \frac{1}{2} \left( \frac{1}{10} \frac{dP_{\rm
isd}^{6F}}{dx_F} + \frac{1}{4} \frac{dP_{\rm isd}^{6C}}{dx_F} \right) +
\frac{1}{10} \frac{dP_{\rm iss}^{6F}}{dx_F} \nonumber
\\ &   & + \, \frac{1}{10} \frac{dP_{\rm isuu}^{8F}}{dx_F} +
 \frac{1}{2} \left( \frac{1}{10} \frac{dP_{\rm
isud}^{8F}}{dx_F} + \frac{1}{7} \frac{dP_{\rm isud}^{8C}}{dx_F} 
\right)\nonumber
\\ &   & + \, \frac{1}{2} \left( \frac{1}{10} \frac{dP_{\rm
isdd}^{8F}}{dx_F} + \frac{2}{7} \frac{dP_{\rm isdd}^{8C}}{dx_F} \right) +
\frac{1}{10} \frac{dP_{\rm issu}^{8F}}{dx_F} 
\nonumber \\ &   & + \, \frac{1}{2} \left( \frac{1}{10} \frac{dP_{\rm
issd}^{8F}}{dx_F} + \frac{2}{12} \frac{dP_{\rm issd}^{8C}}{dx_F} \right) +
 \frac{1}{2} \left( \frac{1}{10} \frac{dP_{\rm
isss}^{8F}}{dx_F} + \frac{3}{16} \frac{dP_{\rm isss}^{8C}}{dx_F} \right)\,\, , 
\label{piprobk0b} 
\\
\frac{dP_{\Lambda}}{dx_F} & = & \frac{dP_{\Sigma^0}}{dx_F} = 
\frac{1}{10} \frac{dP_{\rm
is}^{4F}}{dx_F} + 
 \frac{1}{2} \left( \frac{1}{10} \frac{dP_{\rm
isu}^{6F}}{dx_F} + \frac{1}{4} \frac{dP_{\rm isu}^{6C}}{dx_F} \right) 
+ \frac{1}{10} \frac{dP_{\rm
isd}^{6F}}{dx_F} + \frac{1}{10} \frac{dP_{\rm iss}^{6F}}{dx_F}
\nonumber \\ &   & + \, \frac{1}{2} \left( \frac{1}{10} \frac{dP_{\rm
isuu}^{8F}}{dx_F} + \frac{2}{7} \frac{dP_{\rm isuu}^{8C}}{dx_F} \right) +
 \frac{1}{2} \left( \frac{1}{10} \frac{dP_{\rm
isud}^{8F}}{dx_F} 
+ \frac{2}{7} \frac{dP_{\rm isud}^{8C}}{dx_F} \right)\nonumber
\\ &   & + \, \frac{1}{10} \frac{dP_{\rm
isdd}^{8F}}{dx_F} + 
 \frac{1}{2} \left( \frac{1}{10} \frac{dP_{\rm
issu}^{8F}}{dx_F} + \frac{2}{12} \frac{dP_{\rm issu}^{8C}}{dx_F} \right) 
+ \frac{1}{10} \frac{dP_{\rm
issd}^{8F}}{dx_F} + \frac{1}{10} \frac{dP_{\rm isss}^{8F}}{dx_F} \, \, ,
\label{piproblam} 
\\
\frac{dP_{\Sigma^-}}{dx_F} & = & \frac{1}{10} \frac{dP_{\rm
is}^{4F}}{dx_F} + \frac{1}{10} \frac{dP_{\rm
isu}^{6F}}{dx_F} + \frac{1}{2} \left( \frac{1}{10} \frac{dP_{\rm
isd}^{6F}}{dx_F} + \frac{1}{4} \frac{dP_{\rm isd}^{6C}}{dx_F} \right) +
\frac{1}{10} \frac{dP_{\rm iss}^{6F}}{dx_F} \nonumber
\\ &   & + \, \frac{1}{10} \frac{dP_{\rm isuu}^{8F}}{dx_F} + 
 \frac{1}{2} \left( \frac{1}{10} \frac{dP_{\rm
isud}^{8F}}{dx_F} 
+ \frac{1}{7} \frac{dP_{\rm isud}^{8C}}{dx_F} \right) +
\frac{1}{2} \left( \frac{1}{10} \frac{dP_{\rm
isdd}^{8F}}{dx_F} + \frac{3}{7} \frac{dP_{\rm isdd}^{8C}}{dx_F} \right)
\nonumber
\\ &   & + \, \frac{1}{10} \frac{dP_{\rm issu}^{8F}}{dx_F} 
+ \frac{1}{2} \left( \frac{1}{10} \frac{dP_{\rm
issd}^{8F}}{dx_F} + \frac{2}{12} \frac{dP_{\rm issd}^{8C}}{dx_F} \right) +
\frac{1}{10} \frac{dP_{\rm isss}^{8F}}{dx_F} \, \, ,
\label{piprobsgm} 
\\
\frac{dP_{\Sigma^+}}{dx_F} & = & \frac{1}{10} \frac{dP_{\rm
is}^{4F}}{dx_F} + \frac{1}{10} \frac{dP_{\rm isu}^{6F}}{dx_F} 
+ \frac{1}{10} \frac{dP_{\rm
isd}^{6F}}{dx_F} + \frac{1}{10} \frac{dP_{\rm
iss}^{6F}}{dx_F} + \frac{1}{2} \left( \frac{1}{10} \frac{dP_{\rm
isuu}^{8F}}{dx_F} + \frac{1}{7} \frac{dP_{\rm isuu}^{8C}}{dx_F} \right) 
\nonumber
\\ &   & + \, \frac{1}{10} \frac{dP_{\rm isud}^{8F}}{dx_F} 
+ \frac{1}{10} \frac{dP_{\rm
isdd}^{8F}}{dx_F} + \frac{1}{10} \frac{dP_{\rm issu}^{8F}}{dx_F} 
+\frac{1}{10} \frac{dP_{\rm
issd}^{8F}}{dx_F} + \frac{1}{10} \frac{dP_{\rm isss}^{8F}}{dx_F} \, \, ,
\label{piprobsgp} 
\\
\frac{dP_{\Xi^0}}{dx_F} & = & \frac{1}{10} \frac{dP_{\rm
is}^{4F}}{dx_F} + \frac{1}{10} \frac{dP_{\rm isu}^{6F}}{dx_F} + 
\frac{1}{10} \frac{dP_{\rm
isd}^{6F}}{dx_F} + \frac{1}{10} \frac{dP_{\rm
iss}^{6F}}{dx_F} + \frac{1}{10} \frac{dP_{\rm
isuu}^{8F}}{dx_F} + \frac{1}{10} \frac{dP_{\rm
isud}^{8F}}{dx_F} \nonumber
\\ &   & + \, \frac{1}{10} \frac{dP_{\rm
isdd}^{8F}}{dx_F} + 
 \frac{1}{2} \left( \frac{1}{10} \frac{dP_{\rm
issu}^{8F}}{dx_F} + \frac{1}{12} \frac{dP_{\rm issu}^{8C}}{dx_F} \right) 
+ \frac{1}{10} \frac{dP_{\rm
issd}^{8F}}{dx_F} + \frac{1}{10} \frac{dP_{\rm
isss}^{8F}}{dx_F} \, \, ,
\label{piprobxi0} 
\\
\frac{dP_{\Xi^-}}{dx_F} & = & \frac{1}{10} \frac{dP_{\rm
is}^{4F}}{dx_F} + \frac{1}{10} \frac{dP_{\rm
isu}^{6F}}{dx_F} + \frac{1}{10} \frac{dP_{\rm
isd}^{6F}}{dx_F} + 
 \frac{1}{2} \left( \frac{1}{10} \frac{dP_{\rm
iss}^{6F}}{dx_F} + \frac{1}{7} \frac{dP_{\rm iss}^{6C}}{dx_F} \right) \nonumber
\\ &   & + \, \frac{1}{10} \frac{dP_{\rm
isuu}^{8F}}{dx_F} + \frac{1}{10} \frac{dP_{\rm
isud}^{8F}}{dx_F} + \frac{1}{10} \frac{dP_{\rm
isdd}^{8F}}{dx_F} + 
 \frac{1}{2} \left( \frac{1}{10} \frac{dP_{\rm
issu}^{8F}}{dx_F} + \frac{1}{12} \frac{dP_{\rm issu}^{8C}}{dx_F} \right) 
\nonumber \\ &   & + \, \frac{1}{2} \left( \frac{1}{10} \frac{dP_{\rm
issd}^{8F}}{dx_F} + \frac{2}{12} \frac{dP_{\rm issd}^{8C}}{dx_F} \right) +
 \frac{1}{2} \left( \frac{1}{10} \frac{dP_{\rm
isss}^{8F}}{dx_F} + \frac{3}{16} \frac{dP_{\rm isss}^{8C}}{dx_F} \right)\,\, , 
\label{piprobxim} 
\\
\frac{dP_{\Omega}}{dx_F} & = & \frac{1}{10} \frac{dP_{\rm
is}^{4F}}{dx_F} + \frac{1}{10} \frac{dP_{\rm
isu}^{6F}}{dx_F} + \frac{1}{10} \frac{dP_{\rm
isd}^{6F}}{dx_F} + \frac{1}{10} \frac{dP_{\rm
iss}^{6F}}{dx_F} + \frac{1}{10} \frac{dP_{\rm
isuu}^{8F}}{dx_F} + \frac{1}{10} \frac{dP_{\rm
isud}^{8F}}{dx_F} \nonumber
\\ &   & + \, \frac{1}{10} \frac{dP_{\rm
isdd}^{8F}}{dx_F} + \frac{1}{10} \frac{dP_{\rm
issu}^{8F}}{dx_F} + \frac{1}{10} \frac{dP_{\rm
issd}^{8F}}{dx_F} + \frac{1}{2} \left( \frac{1}{10} \frac{dP_{\rm
isss}^{8F}}{dx_F} + \frac{1}{16} \frac{dP_{\rm isss}^{8C}}{dx_F} \right)\,\, , 
\label{piprobom} 
\\
\frac{dP_{\overline \Lambda}}{dx_F} & = & \frac{dP_{\overline 
\Sigma^0}}{dx_F} = \frac{1}{10} \frac{dP_{\rm
is}^{4F}}{dx_F} + \frac{1}{10} \frac{dP_{\rm
isu}^{6F}}{dx_F} + \frac{1}{2} \left( \frac{1}{10} \frac{dP_{\rm
isd}^{6F}}{dx_F} + \frac{1}{4} \frac{dP_{\rm isd}^{6C}}{dx_F} \right) +
\frac{1}{10} \frac{dP_{\rm iss}^{6F}}{dx_F} \nonumber
\\ &   & + \, \frac{1}{10} \frac{dP_{\rm
isuu}^{8F}}{dx_F} + \frac{1}{2} \left( \frac{1}{10} \frac{dP_{\rm
isud}^{8F}}{dx_F} 
+ \frac{2}{7} \frac{dP_{\rm isud}^{8C}}{dx_F} \right)
+ \frac{1}{2} \left( \frac{1}{10} \frac{dP_{\rm
isdd}^{8F}}{dx_F} + \frac{2}{7} \frac{dP_{\rm isdd}^{8C}}{dx_F} \right) 
\nonumber \\ &   & + \, \frac{1}{10} \frac{dP_{\rm issu}^{8F}}{dx_F} 
+ \frac{1}{2} \left( \frac{1}{10} \frac{dP_{\rm
issd}^{8F}}{dx_F} + \frac{2}{12} \frac{dP_{\rm issd}^{8C}}{dx_F} \right) +
\frac{1}{10} \frac{dP_{\rm isss}^{8F}}{dx_F} \,\, ,
\label{piproblamb} 
\\
\frac{dP_{\overline \Sigma^-}}{dx_F} & = & \frac{1}{10} \frac{dP_{\rm
is}^{4F}}{dx_F} + \frac{1}{10} \frac{dP_{\rm
isu}^{6F}}{dx_F} + \frac{1}{10} \frac{dP_{\rm
isd}^{6F}}{dx_F} + \frac{1}{10} \frac{dP_{\rm
iss}^{6F}}{dx_F} + \frac{1}{10} \frac{dP_{\rm isuu}^{8F}}{dx_F} + 
\frac{1}{10} \frac{dP_{\rm isud}^{8F}}{dx_F} \nonumber
\\ &   & + \, \frac{1}{2} \left( \frac{1}{10} \frac{dP_{\rm
isdd}^{8F}}{dx_F} + \frac{1}{7} \frac{dP_{\rm isdd}^{8C}}{dx_F} \right) +
\frac{1}{10} \frac{dP_{\rm issu}^{8F}}{dx_F} 
+ \frac{1}{10} \frac{dP_{\rm
issd}^{8F}}{dx_F} + \frac{1}{10} \frac{dP_{\rm
isss}^{8F}}{dx_F} \,\, ,
\label{piprobsgmb} 
\\
\frac{dP_{\overline \Sigma^+}}{dx_F} & = & \frac{1}{10} \frac{dP_{\rm
is}^{4F}}{dx_F} +  \frac{1}{2} \left( \frac{1}{10} \frac{dP_{\rm
isu}^{6F}}{dx_F} + \frac{1}{4} \frac{dP_{\rm isu}^{6C}}{dx_F} \right) 
+ \frac{1}{10} \frac{dP_{\rm
isd}^{6F}}{dx_F} + \frac{1}{10} \frac{dP_{\rm
iss}^{6F}}{dx_F} \nonumber
\\ &   & + \, \frac{1}{2} \left( \frac{1}{10} \frac{dP_{\rm
isuu}^{8F}}{dx_F} + \frac{3}{7} \frac{dP_{\rm isuu}^{8C}}{dx_F} \right) +
 \frac{1}{2} \left( \frac{1}{10} \frac{dP_{\rm
isud}^{8F}}{dx_F} 
+ \frac{1}{7} \frac{dP_{\rm isud}^{8C}}{dx_F} \right)\nonumber
\\ &   & + \, \frac{1}{10} \frac{dP_{\rm
isdd}^{8F}}{dx_F} + 
 \frac{1}{2} \left( \frac{1}{10} \frac{dP_{\rm
issu}^{8F}}{dx_F} + \frac{2}{12} \frac{dP_{\rm issu}^{8C}}{dx_F} \right) 
+ \frac{1}{10} \frac{dP_{\rm
issd}^{8F}}{dx_F} + \frac{1}{10} \frac{dP_{\rm
isss}^{8F}}{dx_F} \,\, ,
\label{piprobspb} 
\\
\frac{dP_{\overline \Xi^0}}{dx_F} & = & \frac{1}{10} \frac{dP_{\rm
is}^{4F}}{dx_F} + \frac{1}{10} \frac{dP_{\rm
isu}^{6F}}{dx_F} + \frac{1}{10} \frac{dP_{\rm
isd}^{6F}}{dx_F} + 
 \frac{1}{2} \left( \frac{1}{10} \frac{dP_{\rm
iss}^{6F}}{dx_F} + \frac{1}{7} \frac{dP_{\rm iss}^{6C}}{dx_F} \right) \nonumber
\\ &   & + \, \frac{1}{10} \frac{dP_{\rm
isuu}^{8F}}{dx_F} + \frac{1}{10} \frac{dP_{\rm
isud}^{8F}}{dx_F} + \frac{1}{10} \frac{dP_{\rm
isdd}^{8F}}{dx_F} + \frac{1}{2} \left( \frac{1}{10} \frac{dP_{\rm
issu}^{8F}}{dx_F} + \frac{2}{12} \frac{dP_{\rm issu}^{8C}}{dx_F} \right) 
\nonumber \\ &   & + \, \frac{1}{2} \left( \frac{1}{10} \frac{dP_{\rm
issd}^{8F}}{dx_F} + \frac{1}{12} \frac{dP_{\rm issd}^{8C}}{dx_F} \right) +
 \frac{1}{2} \left( \frac{1}{10} \frac{dP_{\rm
isss}^{8F}}{dx_F} + \frac{3}{16} \frac{dP_{\rm isss}^{8C}}{dx_F} \right)\,\, ,
\label{piprobxi0b} 
\\
\frac{dP_{\overline \Xi^-}}{dx_F} & = & \frac{1}{10} \frac{dP_{\rm
is}^{4F}}{dx_F} + \frac{1}{10} \frac{dP_{\rm
isu}^{6F}}{dx_F} + \frac{1}{10} \frac{dP_{\rm
isd}^{6F}}{dx_F} + \frac{1}{10} \frac{dP_{\rm
iss}^{6F}}{dx_F} + \frac{1}{10} \frac{dP_{\rm
isuu}^{8F}}{dx_F} + \frac{1}{10} \frac{dP_{\rm
isud}^{8F}}{dx_F} \nonumber
\\ &   & + \, \frac{1}{10} \frac{dP_{\rm
isdd}^{8F}}{dx_F} + \frac{1}{10} \frac{dP_{\rm
issu}^{8F}}{dx_F} 
+ \frac{1}{2} \left( \frac{1}{10} \frac{dP_{\rm
issd}^{8F}}{dx_F} + \frac{1}{12} \frac{dP_{\rm issd}^{8C}}{dx_F} \right) +
\frac{1}{10} \frac{dP_{\rm isss}^{8F}}{dx_F} \,\, ,
\label{piprobximb} 
\\
\frac{dP_{\overline \Omega}}{dx_F} & = & \frac{1}{10} \frac{dP_{\rm
is}^{4F}}{dx_F} + \frac{1}{10} \frac{dP_{\rm
isu}^{6F}}{dx_F} + \frac{1}{10} \frac{dP_{\rm
isd}^{6F}}{dx_F} + \frac{1}{10} \frac{dP_{\rm
iss}^{6F}}{dx_F} + \frac{1}{10} \frac{dP_{\rm
isuu}^{8F}}{dx_F} + \frac{1}{10} \frac{dP_{\rm
isud}^{8F}}{dx_F} \nonumber
\\ &   & + \, \frac{1}{10} \frac{dP_{\rm
isdd}^{8F}}{dx_F} + \frac{1}{10} \frac{dP_{\rm
issu}^{8F}}{dx_F} + \frac{1}{10} \frac{dP_{\rm
issd}^{8F}}{dx_F} + \frac{1}{2} \left( \frac{1}{10} \frac{dP_{\rm
isss}^{8F}}{dx_F} + \frac{1}{16} \frac{dP_{\rm isss}^{8C}}{dx_F} \right) 
\, \, .
\label{piprobomb} 
\ee

The strange and antistrange hadron probability distributions from a 
proton projectile are:
\be 
\frac{dP_{K^+}}{dx_F} & = & \frac{1}{2} \left( \frac{1}{10} \frac{dP_{\rm
is}^{5F}}{dx_F} + \frac{2}{4} \frac{dP_{\rm is}^{5C}}{dx_F} \right) + 
 \frac{1}{2} \left( \frac{1}{10} \frac{dP_{\rm
isu}^{7F}}{dx_F} + \frac{3}{5} \frac{dP_{\rm isu}^{7C}}{dx_F} \right) \nonumber
\\ &   & + \, \frac{1}{2} \left( \frac{1}{10} \frac{dP_{\rm
isd}^{7F}}{dx_F} + \frac{2}{5} \frac{dP_{\rm isd}^{7C}}{dx_F} \right) +
 \frac{1}{2} \left( \frac{1}{10} \frac{dP_{\rm iss}^{7F}}{dx_F} + \frac{4}{10} 
\frac{dP_{\rm iss}^{7C}}{dx_F} \right) \nonumber
\\ &   & + \, \frac{1}{2} \left( \frac{1}{10} \frac{dP_{\rm
isuu}^{9F}}{dx_F} + \frac{4}{7} \frac{dP_{\rm isuu}^{9C}}{dx_F} \right) +
 \frac{1}{2} \left( \frac{1}{10} \frac{dP_{\rm
isud}^{9F}}{dx_F} 
+ \frac{3}{7} \frac{dP_{\rm isud}^{9C}}{dx_F} \right)\nonumber
\\ &   & + \, \frac{1}{2} \left( \frac{1}{10} \frac{dP_{\rm
isdd}^{9F}}{dx_F} + \frac{2}{7} \frac{dP_{\rm isdd}^{9C}}{dx_F} \right) +
 \frac{1}{2} \left( \frac{1}{10} \frac{dP_{\rm
issu}^{9F}}{dx_F} + \frac{6}{13} \frac{dP_{\rm issu}^{9C}}{dx_F} \right) 
\nonumber \\ &   & + \, \frac{1}{2} \left( \frac{1}{10} \frac{dP_{\rm
issd}^{9F}}{dx_F} + \frac{4}{13} \frac{dP_{\rm issd}^{9C}}{dx_F} \right) +
 \frac{1}{2} \left( \frac{1}{10} \frac{dP_{\rm
isss}^{9F}}{dx_F} + \frac{6}{19} \frac{dP_{\rm isss}^{9C}}{dx_F} \right)\,\, , 
\label{pprobkp} 
\\
\frac{dP_{K^0}}{dx_F} & = & \frac{1}{2} \left( \frac{1}{10} \frac{dP_{\rm
is}^{5F}}{dx_F} + \frac{1}{4} \frac{dP_{\rm is}^{5C}}{dx_F} \right) + 
 \frac{1}{2} \left( \frac{1}{10} \frac{dP_{\rm
isu}^{7F}}{dx_F} + \frac{1}{5} \frac{dP_{\rm isu}^{7C}}{dx_F} \right) \nonumber
\\ &   & + \, \frac{1}{2} \left( \frac{1}{10} \frac{dP_{\rm
isd}^{7F}}{dx_F} + \frac{2}{5} \frac{dP_{\rm isd}^{7C}}{dx_F} \right) +
 \frac{1}{2} \left( \frac{1}{10} \frac{dP_{\rm
iss}^{7F}}{dx_F} + \frac{2}{10} \frac{dP_{\rm iss}^{7C}}{dx_F} \right) 
\nonumber
\\ &   & + \, \frac{1}{2} \left( \frac{1}{10} \frac{dP_{\rm
isuu}^{9F}}{dx_F} + \frac{1}{7} \frac{dP_{\rm isuu}^{9C}}{dx_F} \right) +
 \frac{1}{2} \left( \frac{1}{10} \frac{dP_{\rm
isud}^{9F}}{dx_F} 
+ \frac{2}{7} \frac{dP_{\rm isud}^{9C}}{dx_F} \right)\nonumber
\\ &   & + \, \frac{1}{2} \left( \frac{1}{10} \frac{dP_{\rm
isdd}^{9F}}{dx_F} + \frac{3}{7} \frac{dP_{\rm isdd}^{9C}}{dx_F} \right) +
 \frac{1}{2} \left( \frac{1}{10} \frac{dP_{\rm
issu}^{9F}}{dx_F} + \frac{2}{13} \frac{dP_{\rm issu}^{9C}}{dx_F} \right) 
\nonumber \\ &   & + \, \frac{1}{2} \left( \frac{1}{10} \frac{dP_{\rm
issd}^{9F}}{dx_F} + \frac{4}{13} \frac{dP_{\rm issd}^{9C}}{dx_F} \right) +
 \frac{1}{2} \left( \frac{1}{10} \frac{dP_{\rm
isss}^{9F}}{dx_F} + \frac{3}{19} \frac{dP_{\rm isss}^{9C}}{dx_F} \right)\,\, , 
\label{pprobk0} 
\\
\frac{dP_{K^-}}{dx_F} & = & \frac{1}{10} \frac{dP_{\rm
is}^{5F}}{dx_F} + 
 \frac{1}{2} \left( \frac{1}{10} \frac{dP_{\rm
isu}^{7F}}{dx_F} + \frac{1}{8} \frac{dP_{\rm isu}^{7C}}{dx_F} \right) 
+ \frac{1}{10} \frac{dP_{\rm
isd}^{7F}}{dx_F} + \frac{1}{10} \frac{dP_{\rm
iss}^{7F}}{dx_F}  \nonumber
\\ &   & + \, \frac{1}{2} \left( \frac{1}{10} \frac{dP_{\rm
isuu}^{9F}}{dx_F} + \frac{2}{13} \frac{dP_{\rm isuu}^{9C}}{dx_F} \right) +
 \frac{1}{2} \left( \frac{1}{10} \frac{dP_{\rm
isud}^{9F}}{dx_F} 
+ \frac{1}{13} \frac{dP_{\rm isud}^{9C}}{dx_F} \right)\nonumber
\\ &   & + \, \frac{1}{10} \frac{dP_{\rm
isdd}^{9F}}{dx_F} +
 \frac{1}{2} \left( \frac{1}{10} \frac{dP_{\rm
issu}^{9F}}{dx_F} + \frac{2}{22} \frac{dP_{\rm issu}^{9C}}{dx_F} \right) 
+ \frac{1}{10} \frac{dP_{\rm
issd}^{9F}}{dx_F} + \frac{1}{10} \frac{dP_{\rm
isss}^{9F}}{dx_F}\,\, ,
\label{pprobkm} 
\\
\frac{dP_{\overline K^0}}{dx_F} & = & \frac{1}{10} \frac{dP_{\rm
is}^{5F}}{dx_F} + \frac{1}{10} \frac{dP_{\rm
isu}^{7F}}{dx_F} + \frac{1}{2} \left( \frac{1}{10} \frac{dP_{\rm
isd}^{7F}}{dx_F} + \frac{1}{8} \frac{dP_{\rm isd}^{7C}}{dx_F} \right) +
\frac{1}{10} \frac{dP_{\rm
iss}^{7F}}{dx_F} \nonumber
\\ &   & + \,  \frac{1}{10} \frac{dP_{\rm
isuu}^{9F}}{dx_F} +
 \frac{1}{2} \left( \frac{1}{10} \frac{dP_{\rm
isud}^{9F}}{dx_F} + \frac{1}{13} \frac{dP_{\rm isud}^{9C}}{dx_F} \right)
+ \frac{1}{2} \left( \frac{1}{10} \frac{dP_{\rm
isdd}^{9F}}{dx_F} + \frac{2}{13} \frac{dP_{\rm isdd}^{9C}}{dx_F} \right) 
\nonumber \\ &   & + \, \frac{1}{10} \frac{dP_{\rm
issu}^{9F}}{dx_F} + \frac{1}{2} \left( \frac{1}{10} \frac{dP_{\rm
issd}^{9F}}{dx_F} + \frac{2}{22} \frac{dP_{\rm issd}^{9C}}{dx_F} \right) +
 \frac{1}{10} \frac{dP_{\rm
isss}^{9F}}{dx_F} \,\, ,
\label{pprobk0b} 
\\
\frac{dP_{\Lambda}}{dx_F} & = & \frac{dP_{\Sigma^0}}{dx_F} =
\frac{1}{2} \left( \frac{1}{10} \frac{dP_{\rm
is}^{5F}}{dx_F} + \frac{2}{4} \frac{dP_{\rm is}^{5C}}{dx_F} \right) + 
 \frac{1}{2} \left( \frac{1}{10} \frac{dP_{\rm
isu}^{7F}}{dx_F} + \frac{3}{8} \frac{dP_{\rm isu}^{7C}}{dx_F} \right) 
\nonumber
\\ &   & + \, \frac{1}{2} \left( \frac{1}{10} \frac{dP_{\rm
isd}^{7F}}{dx_F} + \frac{4}{8} \frac{dP_{\rm isd}^{7C}}{dx_F} \right) +
 \frac{1}{2} \left( \frac{1}{10} \frac{dP_{\rm
iss}^{7F}}{dx_F} + \frac{4}{13} 
\frac{dP_{\rm iss}^{7C}}{dx_F} \right) \nonumber
\\ &   & + \, \frac{1}{2} \left( \frac{1}{10} \frac{dP_{\rm
isuu}^{9F}}{dx_F} + \frac{4}{13} \frac{dP_{\rm isuu}^{9C}}{dx_F} \right) +
 \frac{1}{2} \left( \frac{1}{10} \frac{dP_{\rm
isud}^{9F}}{dx_F} + \frac{6}{13} 
\frac{dP_{\rm isud}^{9C}}{dx_F} \right)\nonumber
\\ &   & + \, \frac{1}{2} \left( \frac{1}{10} \frac{dP_{\rm
isdd}^{9F}}{dx_F} + \frac{6}{13} \frac{dP_{\rm isdd}^{9C}}{dx_F} \right) +
 \frac{1}{2} \left( \frac{1}{10} \frac{dP_{\rm
issu}^{9F}}{dx_F} + \frac{6}{22} \frac{dP_{\rm issu}^{9C}}{dx_F} \right) 
\nonumber \\ &   & + \, \frac{1}{2} \left( \frac{1}{10} \frac{dP_{\rm
issd}^{9F}}{dx_F} + \frac{8}{22} \frac{dP_{\rm issd}^{9C}}{dx_F} \right) +
 \frac{1}{2} \left( \frac{1}{10} \frac{dP_{\rm
isss}^{9F}}{dx_F} + \frac{6}{22} \frac{dP_{\rm isss}^{9C}}{dx_F} \right)\,\, , 
\label{pproblam} 
\\
\frac{dP_{\Sigma^-}}{dx_F} & = & \frac{1}{10} \frac{dP_{\rm is}^{5F}}{dx_F} + 
\frac{1}{10} \frac{dP_{\rm
isu}^{7F}}{dx_F} + \frac{1}{2} \left( \frac{1}{10} \frac{dP_{\rm
isd}^{7F}}{dx_F} + \frac{1}{8} \frac{dP_{\rm isd}^{7C}}{dx_F} \right) +
 \frac{1}{10} \frac{dP_{\rm
iss}^{7F}}{dx_F} \nonumber
\\ &   & + \, \frac{1}{10} \frac{dP_{\rm
isuu}^{9F}}{dx_F} +
 \frac{1}{2} \left( \frac{1}{10} \frac{dP_{\rm
isud}^{9F}}{dx_F} + \frac{1}{13} \frac{dP_{\rm isud}^{9C}}{dx_F} \right)
+ \frac{1}{2} \left( \frac{1}{10} \frac{dP_{\rm
isdd}^{9F}}{dx_F} + \frac{3}{13} \frac{dP_{\rm isdd}^{9C}}{dx_F} \right) 
\nonumber \\ &   & + \, \frac{1}{10} \frac{dP_{\rm
issu}^{9F}}{dx_F} + \frac{1}{2} \left( \frac{1}{10} \frac{dP_{\rm
issd}^{9F}}{dx_F} + \frac{2}{22} \frac{dP_{\rm issd}^{9C}}{dx_F} \right) +
 \frac{1}{10} \frac{dP_{\rm
isss}^{9F}}{dx_F} \,\, ,
\label{pprobsgm} \\
\frac{dP_{\Sigma^+}}{dx_F} & = & \frac{1}{2} \left( \frac{1}{10} \frac{dP_{\rm
is}^{5F}}{dx_F} + \frac{1}{4} \frac{dP_{\rm is}^{5C}}{dx_F} \right) + 
 \frac{1}{2} \left( \frac{1}{10} \frac{dP_{\rm
isu}^{7F}}{dx_F} + \frac{3}{8} \frac{dP_{\rm isu}^{7C}}{dx_F} \right) \nonumber
\\ &   & + \, \frac{1}{2} \left( \frac{1}{10} \frac{dP_{\rm
isd}^{7F}}{dx_F} + \frac{1}{8} \frac{dP_{\rm isd}^{7C}}{dx_F} \right) +
 \frac{1}{2} \left( \frac{1}{10} \frac{dP_{\rm
iss}^{7F}}{dx_F} + \frac{2}{13} 
\frac{dP_{\rm iss}^{7C}}{dx_F} \right) \nonumber
\\ &   & + \, \frac{1}{2} \left( \frac{1}{10} \frac{dP_{\rm
isuu}^{9F}}{dx_F} + \frac{6}{13} \frac{dP_{\rm isuu}^{9C}}{dx_F} \right) +
 \frac{1}{2} \left( \frac{1}{10} \frac{dP_{\rm
isud}^{9F}}{dx_F} + \frac{3}{13} 
\frac{dP_{\rm isud}^{9C}}{dx_F} \right)\nonumber
\\ &   & + \, \frac{1}{2} \left( \frac{1}{10} \frac{dP_{\rm
isdd}^{9F}}{dx_F} + \frac{1}{13} \frac{dP_{\rm isdd}^{9C}}{dx_F} \right) +
 \frac{1}{2} \left( \frac{1}{10} \frac{dP_{\rm
issu}^{9F}}{dx_F} + \frac{6}{22} \frac{dP_{\rm issu}^{9C}}{dx_F} \right) 
\nonumber \\ &   & + \, \frac{1}{2} \left( \frac{1}{10} \frac{dP_{\rm
issd}^{9F}}{dx_F} + \frac{2}{22} \frac{dP_{\rm issd}^{9C}}{dx_F} \right) +
 \frac{1}{2} \left( \frac{1}{10} \frac{dP_{\rm
isss}^{9F}}{dx_F} + \frac{3}{28} \frac{dP_{\rm isss}^{9C}}{dx_F} \right)\,\, , 
\label{pprobsgp} \\
\frac{dP_{\Xi^0}}{dx_F} & = & \frac{1}{10} \frac{dP_{\rm
is}^{5F}}{dx_F} + \frac{1}{10} \frac{dP_{\rm
isu}^{7F}}{dx_F} + \frac{1}{10} \frac{dP_{\rm
isd}^{7F}}{dx_F} +
 \frac{1}{2} \left( \frac{1}{10} \frac{dP_{\rm
iss}^{7F}}{dx_F} + \frac{2}{13} 
\frac{dP_{\rm iss}^{7C}}{dx_F} \right) \nonumber
\\ &   & + \, \frac{1}{10} \frac{dP_{\rm
isuu}^{9F}}{dx_F} + \frac{1}{10} \frac{dP_{\rm
isud}^{9F}}{dx_F} + \frac{1}{10} \frac{dP_{\rm
isdd}^{9F}}{dx_F} +
 \frac{1}{2} \left( \frac{1}{10} \frac{dP_{\rm
issu}^{9F}}{dx_F} + \frac{3}{22} \frac{dP_{\rm issu}^{9C}}{dx_F} \right) 
\nonumber \\ &   & + \, \frac{1}{2} \left( \frac{1}{10} \frac{dP_{\rm
issd}^{9F}}{dx_F} + \frac{2}{22} \frac{dP_{\rm issd}^{9C}}{dx_F} \right) +
 \frac{1}{2} \left( \frac{1}{10} \frac{dP_{\rm
isss}^{9F}}{dx_F} + \frac{6}{28} \frac{dP_{\rm isss}^{9C}}{dx_F} \right)\,\, , 
\label{pprobxi0} \\
\frac{dP_{\Xi^-}}{dx_F} & = & \frac{1}{10} \frac{dP_{\rm
is}^{5F}}{dx_F} + \frac{1}{10} \frac{dP_{\rm
isu}^{7F}}{dx_F} + \frac{1}{10} \frac{dP_{\rm
isd}^{7F}}{dx_F} + \frac{1}{2} \left( \frac{1}{10} \frac{dP_{\rm
iss}^{7F}}{dx_F} + \frac{1}{13} 
\frac{dP_{\rm iss}^{7C}}{dx_F} \right) \nonumber
\\ &   & + \, \frac{1}{10} \frac{dP_{\rm
isuu}^{9F}}{dx_F} + \frac{1}{10} \frac{dP_{\rm
isud}^{9F}}{dx_F} + \frac{1}{10} \frac{dP_{\rm
isdd}^{9F}}{dx_F} +
 \frac{1}{2} \left( \frac{1}{10} \frac{dP_{\rm
issu}^{9F}}{dx_F} + \frac{1}{22} \frac{dP_{\rm issu}^{9C}}{dx_F} \right) 
\nonumber \\ &   & + \, \frac{1}{2} \left( \frac{1}{10} \frac{dP_{\rm
issd}^{9F}}{dx_F} + \frac{2}{22} \frac{dP_{\rm issd}^{9C}}{dx_F} \right) +
 \frac{1}{2} \left( \frac{1}{10} \frac{dP_{\rm
isss}^{9F}}{dx_F} + \frac{3}{28} \frac{dP_{\rm isss}^{9C}}{dx_F} \right)\,\, , 
\label{pprobxim} \\
\frac{dP_{\Omega}}{dx_F} & = & \frac{1}{10} \frac{dP_{\rm
is}^{5F}}{dx_F} + \frac{1}{10} \frac{dP_{\rm
isu}^{7F}}{dx_F} + \frac{1}{10} \frac{dP_{\rm
isd}^{7F}}{dx_F} + \frac{1}{10} \frac{dP_{\rm
iss}^{7F}}{dx_F} + \frac{1}{10} \frac{dP_{\rm
isuu}^{9F}}{dx_F} + \frac{1}{10} \frac{dP_{\rm
isud}^{9F}}{dx_F} \nonumber
\\ &   & + \, \frac{1}{10} \frac{dP_{\rm
isdd}^{9F}}{dx_F} + \frac{1}{10} \frac{dP_{\rm
issu}^{9F}}{dx_F} + \frac{1}{10} \frac{dP_{\rm
issd}^{9F}}{dx_F} + 
 \frac{1}{2} \left( \frac{1}{10} \frac{dP_{\rm
isss}^{9F}}{dx_F} + \frac{1}{28} \frac{dP_{\rm isss}^{9C}}{dx_F} \right)\,\, , 
\label{pprobom} \\
\frac{dP_{\overline \Lambda}}{dx_F} & = & 
\frac{dP_{\overline \Sigma^0}}{dx_F} = \frac{1}{10} 
\frac{dP_{\rm is}^{5F}}{dx_F} + \frac{1}{10} \frac{dP_{\rm
isu}^{7F}}{dx_F} + \frac{1}{10} \frac{dP_{\rm
isd}^{7F}}{dx_F} + \frac{1}{10} \frac{dP_{\rm
iss}^{7F}}{dx_F} \nonumber
\\ &   & + \, \frac{1}{10} \frac{dP_{\rm
isuu}^{9F}}{dx_F} +
 \frac{1}{2} \left( \frac{1}{10} \frac{dP_{\rm
isud}^{9F}}{dx_F} + \frac{1}{7} 
\frac{dP_{\rm isud}^{9C}}{dx_F} \right) + \frac{1}{10} \frac{dP_{\rm
isdd}^{9F}}{dx_F} + \frac{1}{10} \frac{dP_{\rm
issu}^{9F}}{dx_F} 
\nonumber \\ &   & + \, \frac{1}{10} \frac{dP_{\rm
issd}^{9F}}{dx_F} + \frac{1}{10} \frac{dP_{\rm
isss}^{9F}}{dx_F}   \,\, ,
\label{pproblamb} \\
\frac{dP_{\overline \Sigma^-}}{dx_F} & = & \frac{1}{10} \frac{dP_{\rm
is}^{5F}}{dx_F} + \frac{1}{10} \frac{dP_{\rm
isu}^{7F}}{dx_F} + \frac{1}{10} \frac{dP_{\rm
isd}^{7F}}{dx_F} + \frac{1}{10} \frac{dP_{\rm
iss}^{7F}}{dx_F} + \frac{1}{10} \frac{dP_{\rm
isuu}^{9F}}{dx_F} + \frac{1}{10} \frac{dP_{\rm
isud}^{9F}}{dx_F} \nonumber
\\ &   & + \, \frac{1}{2} \left( \frac{1}{10} \frac{dP_{\rm
isdd}^{9F}}{dx_F} + \frac{1}{7} \frac{dP_{\rm isdd}^{9C}}{dx_F} \right) +
 \frac{1}{10} \frac{dP_{\rm
issu}^{9F}}{dx_F} + \frac{1}{10} \frac{dP_{\rm
issd}^{9F}}{dx_F} + \frac{1}{10} \frac{dP_{\rm
isss}^{9F}}{dx_F} \,\, ,
\label{pprobsgmb} \\
\frac{dP_{\overline \Sigma^+}}{dx_F} & = & \frac{1}{10} \frac{dP_{\rm
is}^{5F}}{dx_F} + \frac{1}{10} \frac{dP_{\rm
isu}^{7F}}{dx_F} + \frac{1}{10} \frac{dP_{\rm
isd}^{7F}}{dx_F} + \frac{1}{10} \frac{dP_{\rm
iss}^{7F}}{dx_F} + \frac{1}{2} \left( \frac{1}{10} \frac{dP_{\rm
isuu}^{9F}}{dx_F} + \frac{1}{7} \frac{dP_{\rm isuu}^{9C}}{dx_F} \right) 
\nonumber \\ &   & +
\frac{1}{10} \frac{dP_{\rm isud}^{9F}}{dx_F} + \frac{1}{10} \frac{dP_{\rm
isdd}^{9F}}{dx_F} + \frac{1}{10} \frac{dP_{\rm
issu}^{9F}}{dx_F} + \frac{1}{10} \frac{dP_{\rm
issd}^{9F}}{dx_F} + \frac{1}{10} \frac{dP_{\rm
isss}^{9F}}{dx_F} \,\, ,
\label{pprobsgpb} \\
\frac{dP_{\overline \Xi^0}}{dx_F} & = & \frac{1}{10} \frac{dP_{\rm
is}^{5F}}{dx_F} + \frac{1}{10} \frac{dP_{\rm
isu}^{7F}}{dx_F} + \frac{1}{10} \frac{dP_{\rm
isd}^{7F}}{dx_F} + \frac{1}{10} \frac{dP_{\rm
iss}^{7F}}{dx_F} + \frac{1}{10} \frac{dP_{\rm
isuu}^{9F}}{dx_F} + \frac{1}{10} \frac{dP_{\rm
isud}^{9F}}{dx_F} \nonumber
\\ &   & + \, \frac{1}{10} \frac{dP_{\rm
isdd}^{9F}}{dx_F} +
 \frac{1}{2} \left( \frac{1}{10} \frac{dP_{\rm
issu}^{9F}}{dx_F} + \frac{1}{13} \frac{dP_{\rm issu}^{9C}}{dx_F} \right) 
+ \frac{1}{10} \frac{dP_{\rm
issd}^{9F}}{dx_F} + \frac{1}{10} \frac{dP_{\rm
isss}^{9F}}{dx_F} \,\, ,
\label{pprobxi0b} \\
\frac{dP_{\overline \Xi^-}}{dx_F} & = & \frac{1}{10} \frac{dP_{\rm
is}^{5F}}{dx_F} + \frac{1}{10} \frac{dP_{\rm
isu}^{7F}}{dx_F} + \frac{1}{10} \frac{dP_{\rm
isd}^{7F}}{dx_F} + \frac{1}{10} \frac{dP_{\rm
iss}^{7F}}{dx_F} + \frac{1}{10} \frac{dP_{\rm
isuu}^{9F}}{dx_F} + \frac{1}{10} \frac{dP_{\rm
isud}^{9F}}{dx_F} \nonumber
\\ &   & + \, \frac{1}{10} \frac{dP_{\rm
isdd}^{9F}}{dx_F} 
+ \frac{1}{10} \frac{dP_{\rm
issu}^{9F}}{dx_F} + \frac{1}{2} \left( \frac{1}{10} \frac{dP_{\rm
issd}^{9F}}{dx_F} + \frac{1}{13} \frac{dP_{\rm issd}^{9C}}{dx_F} \right) 
+ \frac{1}{10} \frac{dP_{\rm
isss}^{9F}}{dx_F} \,\, ,
\label{pprobximb} \\
\frac{dP_{\overline \Omega}}{dx_F} & = & \frac{1}{10} \frac{dP_{\rm
is}^{5F}}{dx_F} + \frac{1}{10} \frac{dP_{\rm
isu}^{7F}}{dx_F} + \frac{1}{10} \frac{dP_{\rm
isd}^{7F}}{dx_F} + \frac{1}{10} \frac{dP_{\rm
iss}^{7F}}{dx_F} + \frac{1}{10} \frac{dP_{\rm
isuu}^{9F}}{dx_F} + \frac{1}{10} \frac{dP_{\rm
isud}^{9F}}{dx_F} \nonumber
\\ &   & + \, \frac{1}{10} \frac{dP_{\rm
isdd}^{9F}}{dx_F} + \frac{1}{10} \frac{dP_{\rm
issu}^{9F}}{dx_F} + \frac{1}{10} \frac{dP_{\rm
issd}^{9F}}{dx_F} +
 \frac{1}{2} \left( \frac{1}{10} \frac{dP_{\rm
isss}^{9F}}{dx_F} + \frac{1}{19} \frac{dP_{\rm isss}^{9C}}{dx_F} \right)   
\label{pprobomb} \, \, .
\ee

\end{document}